\documentclass[aip,reprint]{revtex4-1}
\usepackage{graphicx}
\usepackage{dcolumn}
\usepackage{bm}
\usepackage{hyperref}
\usepackage[utf8]{inputenc}
\usepackage[T1]{fontenc}
\usepackage{mathptmx}
\usepackage{etoolbox}
\usepackage{xcolor}
\usepackage{lipsum}
\newcommand{\jb}[1]{\textcolor{black}{#1}} 
\newcommand{\tg}[1]{\textcolor{black}{#1}}
\draft 
\usepackage{float}
\usepackage{comment}
\usepackage{amssymb} 
 \usepackage{amsmath} 
 
\begin{document}

\title{Shear-driven memory effects in carbon black gels}

\author{Julien Bauland}
\affiliation{Univ Lyon, Ens de Lyon, CNRS, Laboratoire de Physique, 69342 Lyon, France}

\author{Thomas Gibaud}
\email[]{Corresponding author, thomas.gibaud@ens-lyon.fr}
\affiliation{Univ Lyon, Ens de Lyon, CNRS, Laboratoire de Physique, 69342 Lyon, France}
\affiliation{Department of Polymer Engineering, IPC, University of Minho, Guimarães, 4804-533 Portugal}

\date{\today}

\begin{abstract}
In recent years, significant effort has been devoted to developing smart materials whose mechanical properties can adapt under physical stimuli. Particulate colloidal gels, which behave as solids but can also flow under stress, have emerged as promising candidates. Resulting from the attractive interaction between their constituents, their network architecture exhibit solid-like properties even at very low volume fractions. This structural flexibility allows them to adopt various configurations and store structural information making them highly susceptible to memory effects. Shear flow, applied through rheometry, offers a simple and effective way to tune their properties and imprint a ``rheological memory'' of the flow history. However, the precise relationship between flow history and viscoelastic response remains elusive, largely due to the limited structural characterization of these systems during flow and after flow cessation. 
Here, we use ultra-small angle X-ray scattering (USAXS) to reveal a strong structural memory in the solid state, where the microstructure formed under shear is retained after flow cessation. We identify two distinct mechanisms of structural memory, as governed by the ratio of viscous to attractive forces, namely, the Mason number. Using recently developed fractal scaling laws, we show that the rheology is fully determined by the gel microstructure. Notably, these gels exhibit a double-fractal architecture, highlighting the remarkably broad range of length scales over which these disordered materials are structured. By clarifying how memory is encoded, our results offer strategies to tune shear sensitivity of colloidal gels and design smart materials.

\end{abstract}

\maketitle 

\section{Introduction}




Drawing inspiration from living matter, an emerging topic in material science is the development of ``smart'' systems whose properties would adapt to an external stimulus~\cite{Jones2016,Jiang2022,Nelson2022}. Owing to the dynamic properties of their microstructure, particulate colloidal gels have become prime candidates for such materials. Experimental studies have shown that their macroscopic properties can be tuned using shear flow~\cite{Koumakis2015,Nelson2022}, ultrasound~\cite{Dages2021}, magnetic~\cite{tasoglu2014} and light~\cite{Chowdhury2021}. 

\tg{Since the seminal work of Koumakis et al.~\cite{Koumakis2015}, the application of shear flow—whether continuous or oscillatory, using a rheometer—has emerged as the most straightforward and accessible method for tuning the properties of shear-rejuvenable colloidal gels and for imprinting a ‘rheological memory’ of their flow history}. The concept of memory refers to the material's ability to retain structural or mechanical changes induced by shear flow, even after the flow is stopped~\cite{Fiocco2014}. This ability arises from the out-of-equilibrium nature and open structure of colloidal gels, which allows for a wide range of distinct configurations, making them particularly susceptible to memory effects.

The interplay between shear flow and the gel's microstructure can be understood through the Mason number ($Mn$), defined as the ratio of the viscous drag force acting on particles to the attractive forces between them~\cite{Mewis2009,Varga2018}. At very low Mason numbers ($Mn \leq 10^{-2}$), the gel behaves as a viscoelastic solid. For intermediate values ($10^{-2} \leq Mn \leq 1$), the constant breakup and reformation of interparticle bonds lead to microstructural rearrangements through which memory is encoded. At high Mason number ($Mn > 1$), the gel is fluidized into a dispersion of particle aggregates, erasing any previously encoded memory. Within this picture, numerical simulations have shown that the structure and strength of colloidal gels can be predicted based on both the Mason number and the timescale of deformation, in a so-called ``time-rate-transformation'' framework~\cite{Jamali2020}. Starting from rest, increasing the shearing time—or total strain—at intermediate Mason numbers leads to a transition from a strong, homogeneous gel network to a phase-separated fluid.

Experimentally, exposition of large flow rates, termed ``rejuvenation'', consistently produces strong and homogeneous gels upon flow cessation~\cite{Koumakis2015}. For depletion gels with a high volume fraction ($\phi = 0.44$), decreasing the rate of shear—applied continuously—lead to heterogeneous weaker gels, with reduced elasticity~\cite{Koumakis2015}. In contrast, when pre-shear is applied under oscillatory motion, the gel elasticity exhibit a minimum at intermediary rate~\cite{Moghimi2017}. Brownian dynamics simulations have shown that gel elasticity negatively correlates with structural heterogeneity, quantified as the void volume in the network. For depletion gels at lower volume fractions ($0.04 \leq \phi \leq 0.1$), further studies revealed a non-monotonic dependence of elasticity on flow magnitude—regardless of the flow type—with strengthening observed at intermediate shear rates~\cite{Das2022}. This behavior is interpreted as over-aging, during which the system is allowed to minimize its free-energy, resulting in higher connectivity.  Consistent with this pictures, the frequency-dependent elasticity of depletion gels was shown to strongly depends on the floppy modes, without change in the average contact number between particles~\cite{Rocklin2021}. 
\tg{In gels where inter-particle interactions arise from van der Waals forces, such as boehmite and carbon black gels, several studies have reported that elasticity increases as flow rate decreases~\cite{Sudreau2022, Dages2022b}}. This increase has been attributed to anisotropic structuring in the shear direction or the interpenetration of fractal clusters, respectively. For silica gels, orthogonal superposition rheometry revealed a ``time-shear'' superposition principle, where shear-mediated changes in aggregate size shift the viscoelastic relaxation time of the gel network~\cite{Colombo2017}.

The above observations show that there is yet no clear picture on the effect of flow on the viscoelasticity of colloidal gels. Beyond the constant risk of experimental artifacts such as sedimentation or heterogeneous flow, a major challenge lies in probing the hierarchical structure of these disordered materials, which spans a wide range of length scales. Numerical simulations have shown that the elasticity of colloidal gels depends on subpopulations of particles, with metrics like mean coordination number insufficient to fully describe it~\cite{Hsiao2012}. A classical approach to linking structure and rheology involves fractal models, where network rigidity depends on the stiffness of intra- or inter-aggregate links~\cite{Shih1990,Wu2001,Bouthier2022b}.  More recently, experimental and numerical studies have shown that gel elasticity stems from minimally interconnected clusters acting as rigid, load-bearing units~\cite{Zaccone2009,Whitaker2019}. Further efforts have focused on identifying these rigid units, for example, through network science approaches~\cite{Nabizadeh2024}.

Given the intimate link between structure and rheology in such material, a deeper understanding of colloidal gel tunability through flow calls for additional structural measurements, both during flow and after flow cessation along with comparisons of the observed structures to rheological properties. In this article, we investigate the shear-induced structuring of carbon black gels at low volume fractions by controlling the shear rate. Using rheometric measurements coupled with small-angle X-ray scattering, we perform time-resolved experiments that probe structural evolution across length scales ranging from 1 to 100 times the size of the primary particles.
Our results show that gels formed after flow cessation retain the structural signature of the shear rate imposed just before flow cessation. We identify two regimes based on the Mason number ($Mn$) imposed just before stopping the flow: (i) a hydrodynamic regime at high $Mn$ where the fractal cluster size in the gel scales directly with shear rate, yielding homogeneous gels.  (ii) An elasto‑plastic regime at low $Mn$ where the clusters densify under shear, producing heterogeneous gels whose microstructure also depends on the duration of the pre‑shear.
Moreover, we find that the gel’s elastic modulus depends non‑monotonically on the pre‑shear rate, reflecting the distinct structural pathways in each regime. These findings provide physical insights into how memory is encoded in the structure of colloidal gels as a function of the shear rate they were exposed to before flow cessation.

\section{Materials and Methods} 
We take carbon black as our primary colloidal particles. Carbon black particles \jb{(Vulcan\textsuperscript{\textregistered}PF, Cabot) are dispersed in mineral oil (RTM17 Rotational Viscometer Standard, Paragon Scientific, $\eta_f = 0.252~\mathrm{Pa \cdot s}$ at $T=25^{\circ}\rm C$)} at volume fractions $\phi_{r_0}$ ranging between 1.2 and 3.2~\% (v/v) as previously reported~\cite{Bauland2024,Bauland2025}. Rheometric measurements are conducted at $T=25^{\circ}\rm C$ using a stress-controlled rheometer (HR20, TA Instrument) equipped with a coaxial cylinder geometry consisting of two polycarbonate cylinders (inner diameter 20~mm, outer diameter 22~mm and height 40~mm). 
Each measurement is preceded by a rejuvenation step, where a shear rate of $\dot{\gamma} = 1000~\rm s^{-1}$ is applied for $60~\rm s$. Following this rejuvenation, the pre-shear protocol involves subjecting the dispersion to the shear rate of interest, $\dot{\gamma}_0$, during $200~\rm s$, after which the flow is stopped to bring the dispersion to rest. \jb{In the rest of the manuscript, $\dot{\gamma}_0$ refers to the amplitude of the shear step from the rejuvenation to the pre-shear rate, while $\dot{\gamma}$ designates the share rate in standard tests.} The gelation and aging processes are subsequently monitored during $400~\mathrm{s}$ by measuring the elastic and viscous moduli under small-angle oscillatory shear with a fixed strain ($\gamma_0 = 0.05~\rm \%$) and frequency ($\omega = 2 \pi~\rm rad/s$). Afterwards, the viscoelastic spectrum of the gel is obtained by varying the frequency at fixed strain ($\gamma_0 = 0.1~\rm \%$). 
The microstructure of the 1.6~\% dispersion is probed during flow at after flow cessation using rheometric tests coupled with ultra-small angle X-ray scattering (USAXS) at the ID02 beamline within the European Synchrotron Radiation Facility (ESRF) in Grenoble, France~\cite{narayanan2020}. The scattering intensity $I(q)$ as function of the scattering wave vector $q$ is derived by subtracting the two-dimensional scattering profile of the mineral oil from that of the CB dispersion. 

\begin{figure*}[t!]
    \includegraphics[scale=0.53, clip=true, trim=0mm 0mm 0mm 0mm]{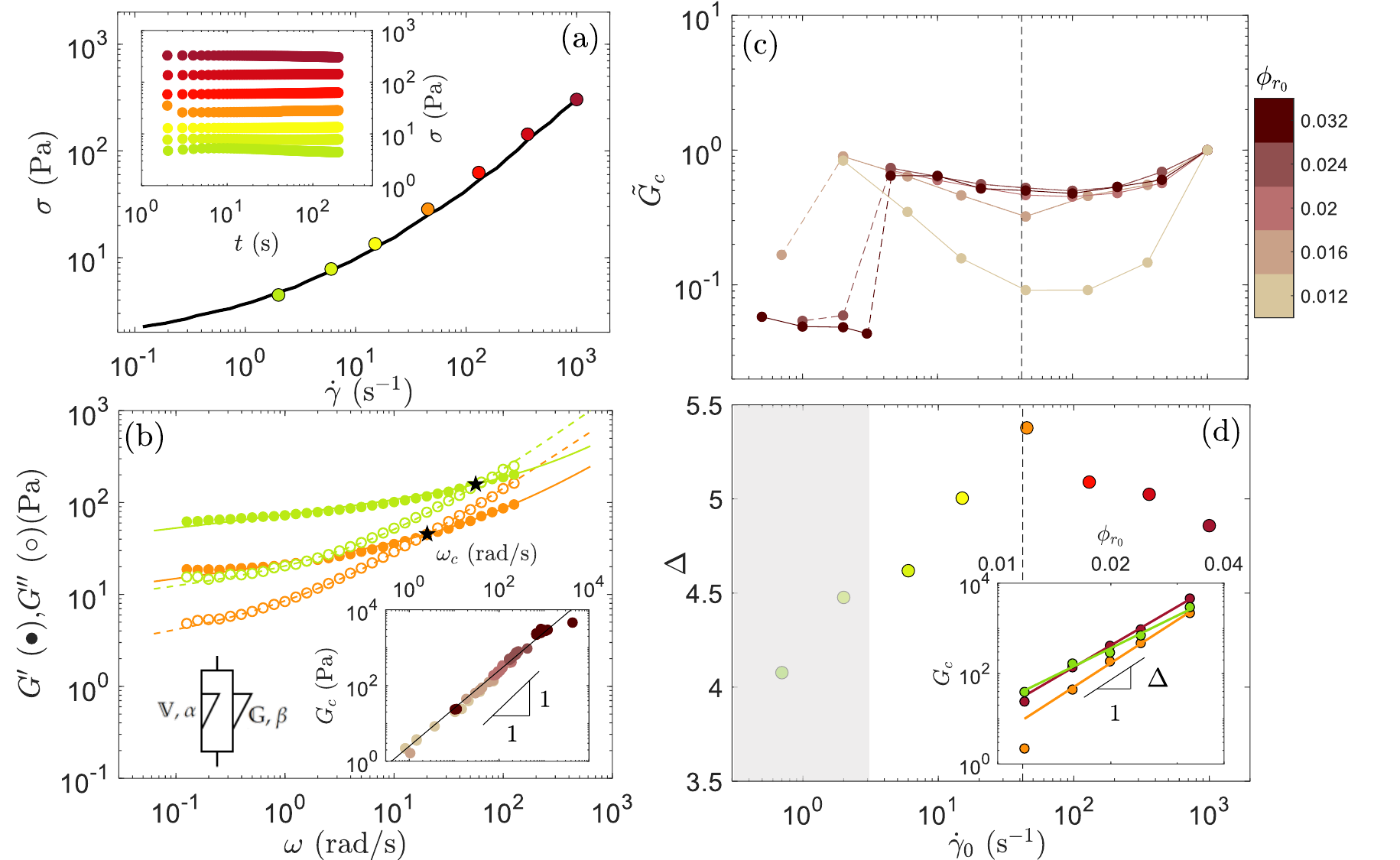}
    \centering
    \caption{(a) Flow curve of the $1.6~\%$ carbon black dispersion measured during a continuous ramping down of the shear rate. Colored markers indicate final stresses from flow step-down experiments at various shear rates (inset). (b) Example of viscoelastic spectra for the $1.6~\%$ dispersion after pre-shearing at $\dot{\gamma} = 45~\rm s^{-1}$ (orange) and $\dot{\gamma} = 1~\rm s^{-1}$ (green). Solid curves represent the best fits using a Kelvin-Voigt fractional model. The crossover point $(G_c, \omega_c)$, where $G^{\prime}_c = G^{\prime\prime}_c$, is highlighted with black markers. The inset shows $G_c$ as a function of $\omega_c$. (c) Rescaled crossover modulus $\tilde{G_c} = G_c(\dot{\gamma}) / G_{c}^p$, where $G_{c}^p$ is the modulus after rejuvenation at $\dot{\gamma} = 10^{3}~\rm s^{-1}$, plotted as a function of the pre-shear rate $\dot{\gamma}$. Color codes for the volume fraction of primary particles $\phi_{r_0}$.   (d) Power-law exponent $\Delta$ from $G_c \propto \phi_{r_0}^{\Delta}$, as a function of pre-shear rate. The vertical line correspond to the critical shear rate $\dot{\gamma^*}=42$~s$^{-1}$ at which $\Delta$ is maximum. The shaded area indicates the antithixotropic regime, where the exponent value (determined for a pre-shear time $t = 200~\rm s$) depends on the shearing time for $t < 10^4~\rm s$.  Inset shows examples of power-law fits.
    }
    \label{fig:rheol}
\end{figure*}

\section{Results}
\subsection{Non monotonic evolution of the gel elasticity with shear rate}
We first investigate the effect of shear on the viscoelastic properties of CB gels by performing step-flow experiments from the rejuvenation rate, followed by flow cessation. For shear rates $\dot{\gamma}_0 > 7~\mathrm{s}^{-1}$, a shearing time of $t = 200~\mathrm{s}$ is applied [inset in Fig.~\ref{fig:rheol}(a)], which is more than sufficient to reach a steady state. In Fig.~\ref{fig:rheol}(a), the final stress measured at the end of the pre-shear step for the 1.6~\% dispersion aligns well with the steady-state flow curve obtained from a continuous flow sweep as reported in~\cite{Hipp2019}. When the shear rate $\dot{\gamma}_0 < 7~\mathrm{s}^{-1}$, however, step-flow experiments involve long transient regimes—referred to as \emph{antithixotropy}~\cite{Bauland2025,Wang2022}—which require extended shearing times and are discussed later in the text.

Fig.~\ref{fig:rheol}(b) shows examples of viscoelastic spectra of CB gels measured after flow cessation (see section \ref{sec:app}.A for the full data set). All the acquired spectra are well described by a Fractional Kelvin-Voigt model [sketched in Fig.~\ref{fig:rheol}(a)], allowing the determination of the crossover point $G^{\prime}(\omega_c) = G^{\prime\prime}(\omega_c) = G_c$, even when this point lies outside the measured frequency range. As in earlier studies, the crossover point ($G_c, \omega_c$)  is used as a metric to describe the gel elasticity independently of the measuring frequency~\cite{Dages2022b}. As reported in \cite{Trappe2000} for a series of concentrations, $G_c$ scales linearly with $\omega_c$ for all tested volume fractions and shear rates. Such linear dependence between $G_c$ and $\omega_c$ is expected for a Kelvin-Voigt model if the viscosity $\eta$ is constant.  Specifically, the characteristic time of the Kelvin-Voigt model, $\tau_c = 1/\omega_c$ follows $\tau_c = G_c/\eta \Leftrightarrow G_c = \eta \omega_c$. This linear scaling confirms that the viscoelastic properties of CB gel can consistently be described by the single parameter $G_c$, that we next refer as the gel ``elasticity''.   

Fig.~\ref{fig:rheol}(c) shows the rescaled crossover modulus ($\tilde{G}_c$) as a function of the pre-shear rate. For each volume fraction, $G_c$ is normalized by the crossover modulus of the gel subjected to the highest pre-shear rate ($\dot{\gamma}_0 = 1000~\rm{s}^{-1}$). First, the influence of the pre-shear rate on the gel's elasticity decreases with increasing volume fraction. Second, the crossover modulus exhibits a minimum at $\dot{\gamma}_0 \approx 40~\rm{s}^{-1}$ for all volume fractions. At this minimum, depending on $\phi_{r_0}$, the modulus is 2 to 10 times lower than that of gels formed at $\dot{\gamma}_0 = 10^3~\rm{s}^{-1}$. 
At very low shear rates, i.e., for $\dot{\gamma}_0 < 7~\rm{s}^{-1}$, the gel elasticity drops by roughly an order of magnitude, which is attributed to antithixotropic restructuring, as discussed later in the text.

We focus here on the regime $\dot{\gamma}_0 > 7~\rm{s}^{-1}$. As previously noted, $G_c$ exhibits a minimum at $\dot{\gamma}_0 \approx 40~\rm{s}^{-1}$ [Fig.~\ref{fig:rheol}(b)]. This non-monotonic behavior is further highlighted by the power-law dependence of $G_c$ on the volume fraction, $G_c \propto \phi_{r_0}^{\Delta}$, for a given pre-shear rate. As shown in Fig.~\ref{fig:rheol}(d), the power-law exponent $\Delta$ reaches a maximum at $\dot{\gamma^*} \approx 40~\rm{s}^{-1}$.

\subsection{Structure of the dispersion before and after flow cessation}
\begin{figure*}[t!]
    \includegraphics[scale=0.53, clip=true, trim=0mm 0mm 0mm 0mm]{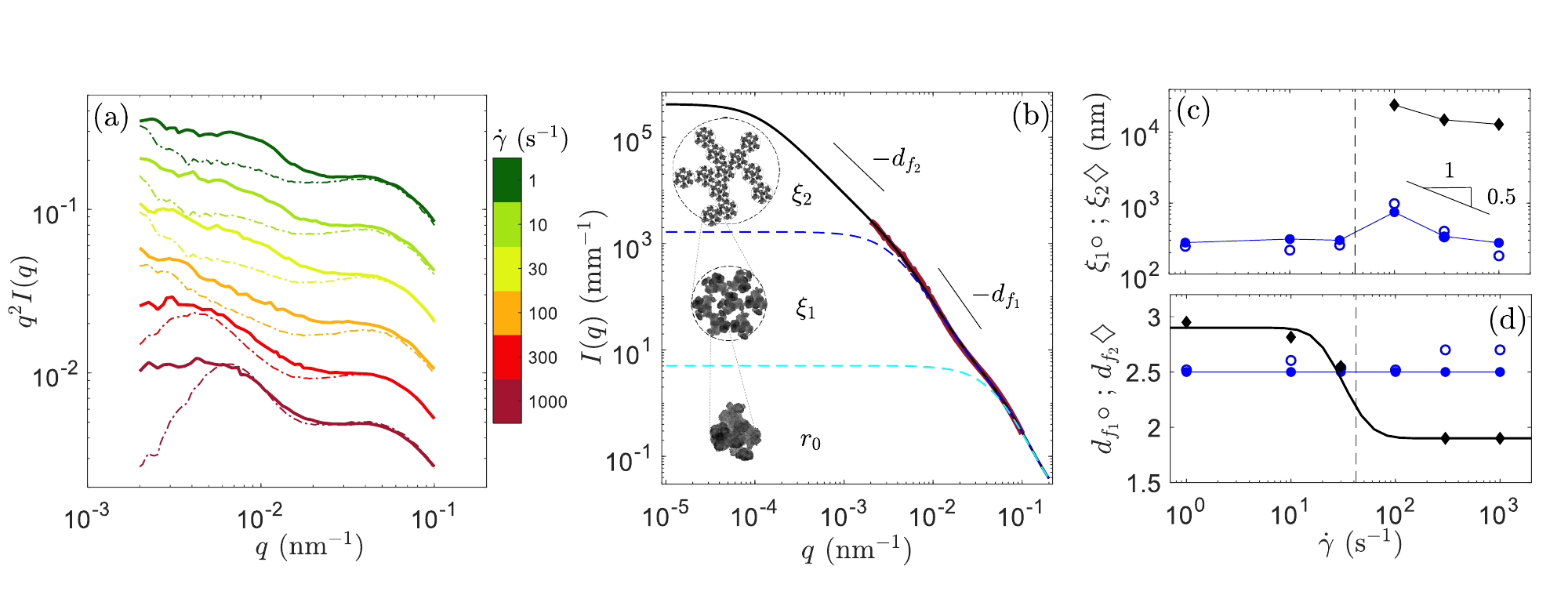}
    \centering
    \caption{(a) Averaged scattering intensity $I(q)$ vs wave vector $q$, measured for the $1.6~\%$ dispersion at various shear rates. Dotted and solid curves represent the structure under flow and after flow cessation, respectively. A horizontal shift and Kratky plots ($q^2 I(q)$ vs $q$) are employed to enhance data visualization. (b) Schematic representation of the hierarchical fractal model used to describe the three structural levels in carbon black dispersions: primary particles ($r_0$), clusters ($\xi_1$), and the network ($\xi_2$). (c)-(d) Dependence of the correlation lengths and fractal dimensions on the shear rate. For the cluster level (blue markers), empty and filled symbols correspond to measurements under flow and after flow cessation, respectively.
    }
    \label{fig:SAXS}
\end{figure*}

To elucidate the microstructural mechanisms underlying the non-monotonic evolution of CB gel elasticity with shear flow magnitude, we investigate the microstructure of the $1.6~\%$ CB dispersion at different shear rates using ultra-small angle X-ray scattering (USAXS). Fig.~\ref{fig:SAXS}(a) presents Kratky plots of the 1D scattering intensity, i.e., $q^2I$ vs.~$q$, measured both during flow (dotted curves) and after flow cessation (solid curves). A break in the slope of the scattering curve $I(q)$ (or a bump in $q^2I(q)$) indicates a characteristic length $\xi$ in the microstructure, with $q \sim 2\pi/\xi$.  Similarly, a fractal organization manifests as a power-law scaling $I \sim q^{-d_f}$, where $d_f$ is the fractal dimension. 

The microstructure of CB dispersions consists of three hierarchical levels~\cite{Dages2022b, Bauland2024}, as depicted in Fig.\ref{fig:SAXS}(b): (i) primary particles $r_0$ (CB particles), which remain intact under shear (see the particles form factor in section~\ref{sec:app}.B), (ii) small clusters of size $\xi_1$ and fractal dimension $d_{f_1}$, formed by the reversible aggregation of the CB particles and (iii) a fractal network of mesh size $\xi_2$ and fractal dimension $d_{f_2}$, formed by the aggregation of the small clusters. In Fig.~\ref{fig:SAXS}(a), the similarity of the scattering curves for a given shear rates and after flow cessation highlights a strong memory of the pre-shear rate in the gel state. 

Focusing on the scattering curves corresponding to the rejuvenation rate ($\dot{\gamma}_0 = 1000~\rm{s}^{-1}$, dark red curves), the scattering spectrum acquired during flow displays two characteristic bumps: one associated with the primary particles $r_0$, centered at $q \sim 8 \times 10^{-2}~\rm{nm}^{-1}$, and another corresponding to the small clusters of size $\xi_1$, centered at $q \sim 7 \times 10^{-3}~\rm{nm}^{-1}$. Upon flow cessation, these small clusters assemble into a percolated network with a mesh size $\xi_2$, which lies outside the experimental window. However, the formation of this network is evidenced by a $q^{-2}$ slope at low $q$ (i.e., a horizontal plateau in the Kratky plot), indicating a fractal dimension $d_{f_2}  \approx 2$. 
Applying a lower pre-shear rate, e.g., $\dot{\gamma}_0 = 300$ or $100~\rm{s}^{-1}$ (light red and orange curves in Fig.~\ref{fig:SAXS}(a)), yields a similar hierarchical structure, but the cluster size $\xi_1$ increases due to weaker shear forces during the pre-shear step, consistent with our previous findings~\cite{Bauland2024}. Notably, at $\dot{\gamma}_0 = 100~\rm{s}^{-1}$, the cluster size $\xi_1$ itself exceeds the measurable range of the setup. Further reducing the pre-shear rate to $\dot{\gamma}_0 \leq 30~\rm{s}^{-1}$ causes $\xi_1$ to return to its original size observed at high shear during rejuvenation, as evidenced by the reappearance of the intermediate bump at $q \sim 8 \times 10^{-2}~\rm{nm}^{-1}$. At such low shear rates, clusters $\xi_1$ are already structured into larger agglomerates during flow, as indicated by the power-law scattering at low $q$. After flow cessation, the exponent of the low-$q$ power-law scattering exceeds 2, indicating that the resulting agglomerates are denser than the fractal networks formed at higher shear rates.

To quantitatively describe these structural changes, the scattering curves are fitted using a hierarchical fractal model~\cite{Hipp2021}, in which each structural level is described by a mass fractal structure factor [Fig.~\ref{fig:SAXS}(b)]. The resulting fitting parameters are shown in Fig.~\ref{fig:SAXS}(c) (see section \ref{sec:app}.B for a detailed description of the fitting procedure). Focusing on the highest shear rates, $\dot{\gamma}_0 = 1000$, $300$, and $100~\mathrm{s}^{-1}$, the cluster size $\xi_1$ increases with decreasing $\dot{\gamma}_0$, following a scaling law $\xi_1 \propto \dot{\gamma}_0^{-0.5}$, both during flow \tg{, as reported in \cite{Bauland2024,Varga2018},} and after flow cessation, as indicated by the empty and solid blue markers in Fig.~\ref{fig:SAXS}(c). In this regime, the cluster size reflects an equilibrium length scale governed by the competition between particle aggregation and erosion~\cite{Bouthier2023}. The fractal dimension of the clusters, however, remains constant at $d_{f_1} = 2.5$ [Fig.~\ref{fig:SAXS}(d)], in agreement with our previous observations~\cite{Bauland2024}. Upon flow cessation, these small clusters assemble into a network with mesh size $\xi_2$ and a fractal dimension $d_{f_2} = 2$, consistent with the reaction-limited aggregation of initially flowing particles~\cite{Lazzari2016}. \tg{Since $\xi_2$ exceeds the length scales accessible by SAXS, assuming the network is homogeneous, we estimate $\xi_2$ based on mass conservation of primary particles $r_0$:} 
\begin{equation}
    \rho = \frac{\bigg(\frac{\xi_2}{\xi_1}\bigg)^{d_{f2}} \bigg(\frac{\xi_1}{r_0}\bigg)^{d_{f1}}}{\xi_2^3}. \\
\end{equation}
\tg{The number density, $\rho$, is obtained from the CB concentration, while the parameters 
$\xi_1$, $d_{f1}$, $d_{f2}$, and $r_0$ are extracted from the SAXS fits 
(see Section~\ref{sec:app}.B for further details). } 
We find that $\xi_2$ is on the order of tens of microns at high shear rates [black markers in Fig.~\ref{fig:SAXS}(c)]. In Fig.~\ref{fig:SAXS}(b), the solid black line represents the extrapolated fit for $\dot{\gamma}_0 = 1000~\mathrm{s}^{-1}$ (after flow cessation), with $\xi_2 = 10~\mathrm{\mu m}$ imposed by the mass conservation of primary particles. 
In summary, at high shear rates ($\dot{\gamma}_0 \geq 100~\mathrm{s}^{-1}$), the size of the cluster $\xi_1$ depends on the pre-shear rate, while their fractal dimension remains constant at $d_{f_1} = 2.5$. Upon flow cessation, clusters consistently assemble into an open network characterized by a fractal dimension $d_{f_2} = 2$.

Further reducing the shear rate, to $\dot{\gamma}_0 \leq 40~\mathrm{s}^{-1}$, the cluster size $\xi_1$ matches the value measured at the rejuvenation rate [Fig.~\ref{fig:SAXS}(c)]. In other words, the shear rate imposed before flow cessation is so low that it no longer affects the small-scale structure $\xi_1$ formed during rejuvenation at 1000~s$^{-1}$. At these low shear rates, the ratio of viscous shear forces to attractive inter-particle forces—quantified by the Mason number—is less than unity. This leads to the assembly of clusters $\xi_1$ into larger and denser structures. Indeed, after flow cessation, the fractal dimension of the network, $d_{f_2}$, increases to 2.6 at $\dot{\gamma}_0 = 30~\mathrm{s}^{-1}$ and up to 2.9 at $\dot{\gamma}_0 = 1~\mathrm{s}^{-1}$ [black markers in Fig.~\ref{fig:SAXS}(d)]. Concomitant with this densification, \tg{the estimation of }the mesh size $\xi_2$, calculated from mass conservation, exceeds the rheometer gap size (see Fig.~\ref{fig:supSAXS3} in the Appendix). This unphysical result suggests that the structure is no longer a homogeneous network characterized by a single mesh size $\xi_2$, but is instead further organized at larger length scales that lie beyond the resolution of USAXS. In other words, to satisfy mass conservation of the primary particles, \tg{large} voids must exist at high length scales.

\begin{figure}[t!]
    \includegraphics[scale=0.75, clip=true, trim=0mm 0mm 0mm 0mm]{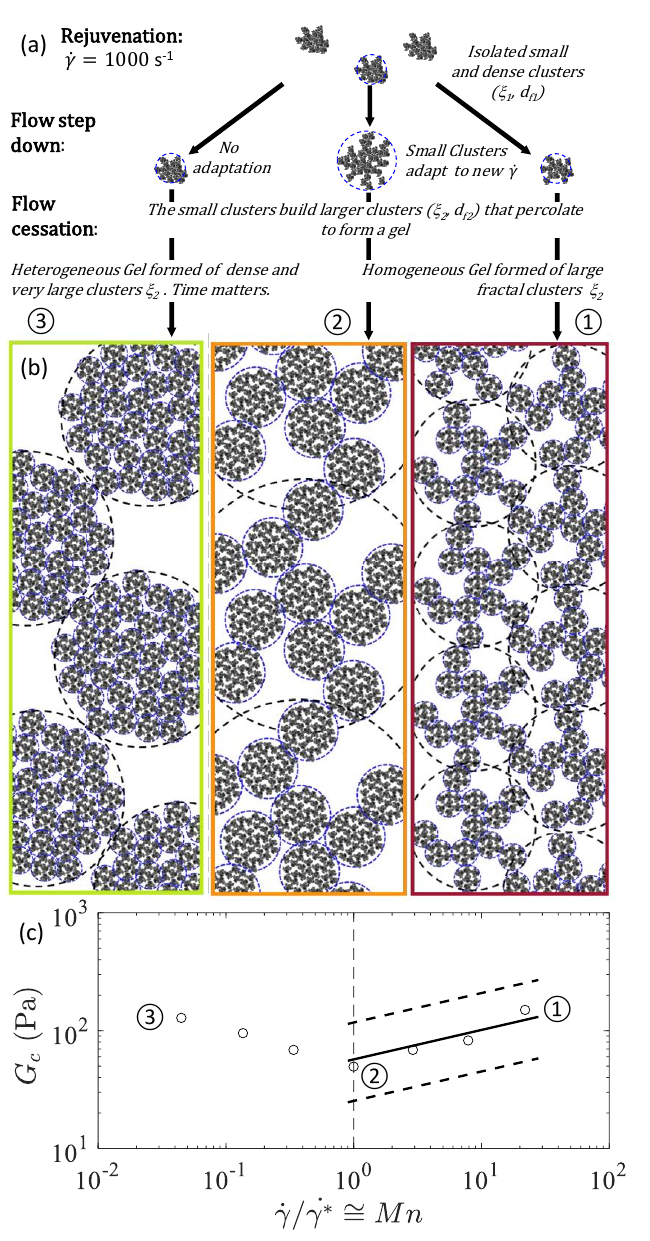}
    \centering
    \caption{Interplay between the gel structure and its mechanical properties. (a) Schematic illustrating that shear history shape the hierarchical organization of the  clusters. (b) The schematic depicts the microstructure of the CB gel, composed of clusters $\xi_1$ (blue dotted lines) assembled into agglomerates $\xi_2$ (black dotted lines). From right to left: \textcircled{1} $Mn \gg 1$, homogeneous network of mesh size $\xi_2$ composed of small clusters $\xi_1$; \textcircled{2} $Mn \sim 1$, homogeneous network of mesh size $\xi_2$ composed of large clusters $\xi_1$; \textcircled{3} $Mn \ll 1$, heterogeneous network of dense mesh size $\xi_2$ composed of small clusters $\xi_1$. (c) Crossover modulus $G_c$ of the $1.6~\%$ dispersion as a function of $\dot{\gamma}/\dot{\gamma^*}$, where $\dot{\gamma^*} = 40~\mathrm{s}^{-1}$ is the critical shear rate determined from Fig.~\ref{fig:rheol}. In the text, we discuss that this normalized shear rate can be identified with the Mason number $Mn$. The solid line represents $G_c^{model}$ (Eq~\ref{eq:G}), with structural parameters obtained from USAXS and ($U = 38~k_B T$,$\delta = 4.4~\mathrm{nm}$). Dash line represent $G_c^{model}$ by varying ($U$, $\delta$) by $\pm$ 25\%.
    }
    \label{fig:model}
\end{figure}

\section{Discussion}

\tg{The microstructural scenario revealed by the Rheo-USAXS data shows two distinct regimes at low and high shear rates separated by  a critical shear rate $\dot{\gamma^*}$ comprise between 30 and 100~$s^{-1}$. To rationalize these findings, we explore two complementary approaches: calculating the Mason number to determine the threshold shear rate that separates the two regimes and modeling the elastic moduli in the high shear rate regime}.
\tg{These two approaches require, as input, a reliable estimate of the interaction parameters between the particles and the selection of an appropriate length scale for the drag force.}
\tg{The properties of the gel obtained at low shear rates is analyzed within the framework of the antithixotropic regime.}

\subsection{Mason Number, $Mn$, as a metric for shear memory}
\label{section:mn}
\tg{We first examine the Mason number, defined as the ratio of drag to interaction forces, as a metric for the shear memory effect in CB gels:}
\begin{equation}
Mn = \frac{6 \eta_f R^2 \dot{\gamma}}{U / \delta}
\label{eq:Mn}
\end{equation}
\tg{where $\eta_f$ is the solvent viscosity, $U$ and $\delta$ the depth of the attraction between particle of size $R$ and the range of the attraction.}
\tg{We use the length scale $R = R_{g1}=430$~nm, the cluster radius of gyration at the rejuvenation rate (1000~s$^{-1}$), as the relevant hydrodynamic length scale. This choice is motivated by the fact that, during a step-down in shear, the viscous drag acts on pre-existing clusters rather than on individual primary particles. These rejuvenated clusters are the smallest stable structural units formed under high shear, which persist and reorganize as the shear rate decreases. 
We then take that the inter-cluster attraction potential is equal to that of the CB particles.}
\tg{We determine $U$ and $\delta$ by mapping the CB van der Waal interaction potential~\cite{dagastine2002} onto a square-well potential of corresponding depth and range and optimizing the value of $U$ based on our experimental data (see Appendix~\ref{sec:app}C for details of the calculation).}
\tg{Such an optimization leads to $Mn = 1$ at $\dot{\gamma}^* = 40~\mathrm{s}^{-1}$ for $U = 38~k_B T$ and $\delta = 4.4~\mathrm{nm}$. 
The value of $U$ is approximately 25\% higher than that reported in the literature~\cite{Varga2019, Trappe2007}, which was obtained for a different type of CB particle dispersed in a different mineral oil. 
Within this margin of error, we therefore establish that $\dot{\gamma} / \dot{\gamma}^*$ can be identified with $Mn$, i.e., $\dot{\gamma} / \dot{\gamma}^* \cong Mn$
}


\tg{As illustrated in the schematics in Fig.~\ref{fig:model} (a-b), starting from the rejuvenation state, performing a small flow step-down such that $Mn \gg 1$, the equilibrium cluster size is governed by the viscous stress imposed during the pre-shear at $\dot{\gamma_0}$. After flow cessation, these clusters assemble into a homogeneous network with mesh size $\xi_2$. Conversely, larger flow step-downs lead to $Mn<1$. In this regime, attractive forces dominate over viscous forces, promoting the formation of a heterogeneous network composed of dense, very large clusters. 
This behavior underlies the antithixotropic nature of the CB dispersion at low shear rates~\cite{Wang2022a, Bauland2025}, as discussed in detail later in the article.
}

\subsection{Gel elasticity}
\tg{Let us then examine the evolution of the elastic moduli, limiting our analysis to the homogeneous gel obtained in the high shear rate regime ($Mn>1$). In the viscous regime decreasing the pre-shear rate increases the size of the building blocks $\xi_1$. From fractal theory, it is well established that the elasticity of a fractal floc scales inversely with its size~\cite{Shih1990}. Fully capturing the three-level structural hierarchy evidenced by USAXS and mass conservation, we use the experimentally determined parameters to model the decrease in gel elasticity using a cluster-of-clusters model~\cite{Bouthier2022}, which accounts for the double fractal nature of CB dispersions. In the stretching-dominated weak-link regime, where elasticity arises from the deformation of inter-cluster links, the gel elasticity is modeled by:}

\begin{equation}
    G_c^{model} = \frac{U}{a\delta^2}\phi\bigg(\frac{\xi_1}{\xi_2}\bigg)^{d_{f_2}-2} \bigg(\frac{r_0}{\xi_1}\bigg)^{d_{f_1}-2}  .
    \label{eq:G}
 \end{equation}


\noindent\tg{In Fig.~\ref{fig:model}(c)), we show that, using the SAXS structural parameters from Fig.~\ref{fig:SAXS}(c--d), the model in Eq.~\ref{eq:G} reproduces the scaling of $G_c$ with shear rate remarkably well, thereby validating the model and the origin of the elasticity in such gels. 
The absolute magnitude of $G_c^{\mathrm{model}}$ in Eq.~\ref{eq:G} is set by the prefactor $\frac{U}{a \delta^2} \phi$, and unsurprisingly, it matches the experimental values because $U = 38~k_B T$ and $\delta = 4.4~\mathrm{nm}$ were optimized accordingly, as detailed in Appendix~\ref{sec:app}. The fact that both $\dot{\gamma}^*$ and $G_c$ can be fitted using the same pair of parameters ($U$, $\delta$), in reasonable agreement with literature values, gives us confidence in their accuracy. 
}


For low shear rates, the relationship between elasticity and structure becomes less straightforward. In this regime, which is dominated by elastic contributions to the stress, rheological modeling is not possible due to the emergence of higher-order structural organization, as indicated by the mass conservation analysis. While the cluster size $\xi_1$ remains inherited from the rejuvenation state, these clusters assemble into dense agglomerates, likely reducing both the effective volume fraction and the coordination number among agglomerates. The coordination number between particles has previously been identified as a key factor governing the elasticity of colloidal gels~\cite{Zaccone2009a, Zaccone2011}. This perspective aligns with the observed attenuation of pre-shear effects on gel elasticity as the volume fraction increases, as shown in Fig.~\ref{fig:rheol}(c). The minimum in elasticity for $\dot{\gamma}/\dot{\gamma^*}= 1$ (vertical dotted line) decreases between $\phi_{r_0} = 1.2$ and $2~\%$, after which it plateaus. 

Assuming that the cluster size and fractal dimension under rejuvenation rate are independent of volume fraction~\cite{Bauland2024}, the effective volume fraction of clusters can be estimated as $\phi_{\xi_1} \approx \phi_{r_0} (\xi_1 / r_0)^{3 - d_{f_1}}$, yielding $\phi_{\xi_1} = 6.5~\%$ for $\phi_{r_0} = 2~\%$. This value marks an upper bound beyond which CB gels become moderately sensitive to shear in the intermediate shear-rate regime. Further decreasing the shear rate at a fixed volume fraction results in an increase in elasticity [Fig.~\ref{fig:model}], likely due to an ``overaging'' of the heterogeneous network, where complex microstructural evolution drives the system into deeper energy minima under shear~\cite{Moghimi2017}.

\subsection{Antithixotropic regime in the limit $Mn<1$}
\begin{figure}[t!]
    \includegraphics[scale=0.55, clip=true, trim=0mm 0mm 0mm 0mm]{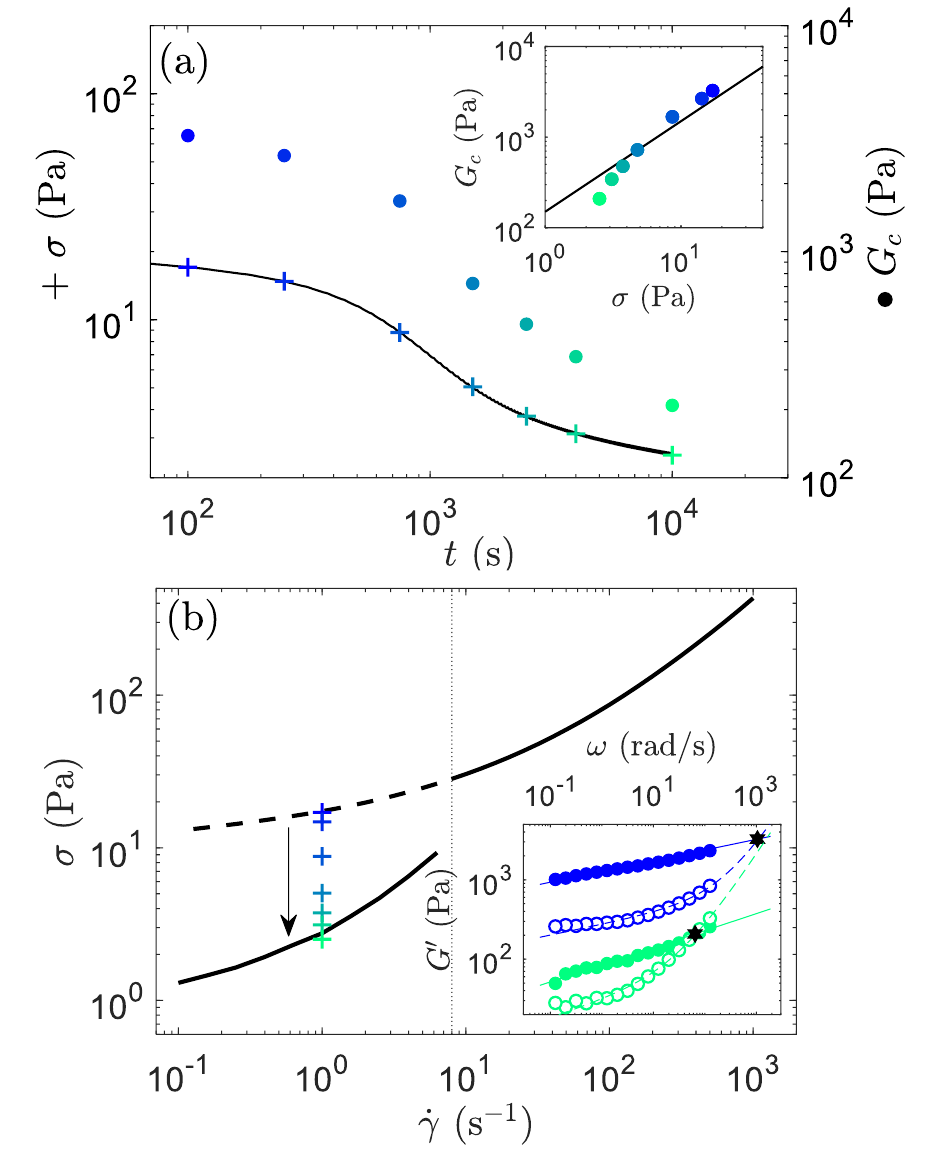}
    \centering
    \caption{(a) Effect of shearing time at $\dot{\gamma} = 1~\mathrm{s}^{-1}$ on the shear stress (crosses) and gel elasticity (dots) after flow cessation for the $3.2~\%$ dispersion. The inset shows the crossover modulus $G_c$ as a function of the shear stress measured just before flow cessation. The black line represents the best linear fit. (b) Flow curve of the $3.2~\%$ CB dispersion. Solid lines correspond to the steady-state flow curve, while the dotted line represents the transient flow curve obtained during a fast flow sweep. The discontinuity at $\dot{\gamma} \approx 10~\mathrm{s}^{-1}$ marks the upper limit of the antithixotropic regime. Markers indicate the time evolution of the stress shown in (a), using the same color code. Inset displays example of viscoelastic spectra.
    }
    \label{fig:anti}
\end{figure}

While dispersions with higher particle volume fractions are less sensitive to pre-shear at intermediate Mason numbers, they exhibit a pronounced reduction in elasticity when exposed over time to very low Mason numbers ($Mn \ll 1$) [Fig.~\ref{fig:rheol}(c)].  
This dramatic decrease is attributed to antithixotropic restructuring of the CB dispersion, during which the fractal nature of the network is lost. The resulting structure consists of large, dense, and loosely connected agglomerates~\cite{Bauland2025}. We have previously shown that the occurrence and timescale of this phenomenon depend on the particle volume fraction, but consistently occur below a critical shear rate $\dot{\gamma} \approx 7~\mathrm{s}^{-1}$ where elastic contributions to the stress become dominant. As shown in Fig.~\ref{fig:anti}, antithixotropy manifests as a slow decay in stress over time when a constant shear rate is applied. Reaching a steady state at $\dot{\gamma} = 1~\mathrm{s}^{-1}$ requires prolonged shearing, on the order of $t \approx 10^4~\mathrm{s}$. As a result, the steady-state flow curve [solid line in Fig.~\ref{fig:anti}(b)] displays a discontinuity at $\dot{\gamma} \approx 7~\mathrm{s}^{-1}$, marking the onset of the antithixotropic regime. The time evolution of the stress, colored identically to Fig.~\ref{fig:anti}(a), demonstrates that the flow properties of these dispersions can be tuned at will through the shearing time~\cite{Ovarlez2013a}.

To illustrate the effect of shearing time in the antithixotropic regime~\cite{Bauland2025,Wang2022} on the resulting gel elasticity for the $3.2~\%$ dispersion, we perform flow cessation at various times along the stress decay observed at $\dot{\gamma}=1\ \mathrm{s}^{-1}$ [see inset in Fig.~\ref{fig:anti}(a)] and determine the resulting gel elasticity $G_c$ after flow cessation (right axis). The decrease in $G_c$ with shearing time closely follows the reduction in shear stress. 
Gels formed after 100~s of pre‑shear at $\dot{\gamma}=1\ \mathrm{s}^{-1}$ exhibit an elastic modulus $G_c \simeq 3000$~Pa, whereas extending the pre‑shear to $10^4\,$s under the same shear rate reduces  the gel elastic modulus to $G_c \simeq 200$~Pa. We note, as shown in Fig.~\ref{fig:anti}(b)-inset that the viscolelastic spectrum of the gel in antithixotropic regime sill can be fitted by the same fractional Kelvin–Voigt model  as in Fig.~\ref{fig:rheol}(b) and still display $G_c = \eta \omega_c$.
Moreover, we have previously shown that the steady-state reached after antithixotropy leads to a ``partial memory loss'' of the flow history as it results in a unique structure that is independent of the pre-shear rate, provided $\dot{\gamma}_0 \approx 10~\mathrm{s}^{-1}$. In Fig.~\ref{fig:rheol}(c), focusing on the $3.2~\%$ dispersion (dark red markers), the gel elasticity indeed remains constant for $0.3 \leq \dot{\gamma}_0 \leq 3~\mathrm{s}^{-1}$. Thus, while the elasticity of dilute suspensions can be tuned at high and intermediate Mason numbers, the elasticity of more concentrated dispersions can be controlled through shearing at very low Mason numbers, owing to the antithixotropic restructuring that occurs under these conditions.


\section{Conclusion}

In this work, we investigated the shear-driven memory of colloidal gels by performing structural measurements both during flow and after flow cessation. Our USAXS experiments first reveal a strong structural memory of the structures formed before flow cessation in the gel state, in particular we showed that the elasticity shows a minimum around a critical preshear rate $\dot{\gamma^*}$. Second, we identify two distinct mechanisms by which rheological memory is encoded. At high shear rate, memory is stored through the size of gel building blocks, namely, small clusters of primary particles, whose equilibrium size is set by the viscous stress. In this regime, the structure of gels formed after flow cessation consists of a homogeneous network. 
At low shear rate, the duration of pre-shear prior to flow cessation strongly influences the gel structure due to antithixotropic behavior. On short timescales, memory is encoded through elastic stress–driven densification of the mesoscopic fractal network, leading to a heterogeneous structure. However, at very low shear rate, high particle volume fractions, and prolonged shearing, a ``partial memory loss'' occurs: shear-driven compaction becomes so pronounced that the original fractal network is fully erased, yielding a loosely connected assembly of large, dense clusters. \tg{Third, we suggest that this scenario can be rationalized in terms of the Mason number, i.e., the balance between viscous shear forces and attractive interparticle forces, by identifying $Mn=\dot{\gamma}/\dot{\gamma^*}$}. The calculation of $Mn$ lead us to identify the clusters formed under the highest shear rate (rejuvenation rate) as the relevant structural entities subjected to the drag force for Mason number calculations. This insight underscores that rheological memory, in this system, cannot be entirely erased by shear, leading to a complex interplay between gel mechanics and shear history. Notably, a different structural evolution is expected during a flow sweep—where clusters can gradually adapt—compared to the abrupt step-down in shear rate performed here.

From a practical perspective, this work paves the way for designing colloidal gels with controlled structure and mechanical properties without altering formulation or volume fraction.
\tg{This could have implications in additive manufacturing where external fields such as an additional shear coupled with 3D printing allow tuning the microstructure and the properties of the printed materials~\cite{raney2018}.}
We demonstrate that the gel elasticity after flow cessation is governed by the Mason number during pre-shear and the resulting microstructure. Notably, while gels with higher volume fractions appear less tunable at intermediate $Mn$, as the microstructure has fewer degrees of freedom, their elasticity can be drastically altered at low $Mn$ thanks to antithixotropy.


Finally, our structural measurements reveal that this disordered material self-organizes across an exceptionally broad range of length scales, spanning over two orders of magnitude beyond the primary particle size. This feature is likely a key factor in enabling a strong sensitivity to shear memory. 
Our estimation of the mesh size in the double-fractal network shows that it extends into the tens-of-microns range. To fully elucidate the link between structure and rheology, future work should incorporate mesoscopic-scale structural measurements, using techniques such as X-ray tomography~\cite{Link2023} or light-sheet fluorescence microscopy~\cite{Spicer2025} for transparent systems.

\section*{Conflicts of interest}
There are no conflicts to declare.

\section*{Data availability} 
The data that support the findings of this study are available from the corresponding author upon reasonable request.

\section*{Acknowledgements}
The authors are especially grateful to the ESRF for beamtime at the beamline ID02 (proposal SC-5236) and Theyencheri Narayanan for the discussions and technical support during the USAXS measurements. The authors also acknowledge fruitful discussions with T. Divoux, A. Poulesquen, S. Manneville.
This work was supported by the Région Auvergne-Rhône-Alpes ``Pack Ambition Recherche'', the LABEX iMUST (ANR-10-LABX-0064) of Université de Lyon, within the program ``Investissements d'Avenir'' (ANR-11-IDEX-0007), the ANR grants (ANR-18-CE06-0013, ANR-21-CE06-0020-01, ANR-TrainGel) and European Union’s Horizon Europe Framework Programme (HORIZON) under the Marie Skłodowska-Curie Grant Agreement 101120301. This work benefited from meetings within the French working group GDR CNRS 2019 ``Solliciter LA Matière Molle'' (SLAMM).


\begin{thebibliography}{47}%
\makeatletter
\providecommand \@ifxundefined [1]{%
 \@ifx{#1\undefined}
}%
\providecommand \@ifnum [1]{%
 \ifnum #1\expandafter \@firstoftwo
 \else \expandafter \@secondoftwo
 \fi
}%
\providecommand \@ifx [1]{%
 \ifx #1\expandafter \@firstoftwo
 \else \expandafter \@secondoftwo
 \fi
}%
\providecommand \natexlab [1]{#1}%
\providecommand \enquote  [1]{``#1''}%
\providecommand \bibnamefont  [1]{#1}%
\providecommand \bibfnamefont [1]{#1}%
\providecommand \citenamefont [1]{#1}%
\providecommand \href@noop [0]{\@secondoftwo}%
\providecommand \href [0]{\begingroup \@sanitize@url \@href}%
\providecommand \@href[1]{\@@startlink{#1}\@@href}%
\providecommand \@@href[1]{\endgroup#1\@@endlink}%
\providecommand \@sanitize@url [0]{\catcode `\\12\catcode `\$12\catcode `\&12\catcode `\#12\catcode `\^12\catcode `\_12\catcode `\%12\relax}%
\providecommand \@@startlink[1]{}%
\providecommand \@@endlink[0]{}%
\providecommand \url  [0]{\begingroup\@sanitize@url \@url }%
\providecommand \@url [1]{\endgroup\@href {#1}{\urlprefix }}%
\providecommand \urlprefix  [0]{URL }%
\providecommand \Eprint [0]{\href }%
\providecommand \doibase [0]{http://dx.doi.org/}%
\providecommand \selectlanguage [0]{\@gobble}%
\providecommand \bibinfo  [0]{\@secondoftwo}%
\providecommand \bibfield  [0]{\@secondoftwo}%
\providecommand \translation [1]{[#1]}%
\providecommand \BibitemOpen [0]{}%
\providecommand \bibitemStop [0]{}%
\providecommand \bibitemNoStop [0]{.\EOS\space}%
\providecommand \EOS [0]{\spacefactor3000\relax}%
\providecommand \BibitemShut  [1]{\csname bibitem#1\endcsname}%
\let\auto@bib@innerbib\@empty
\bibitem [{\citenamefont {Jones}\ and\ \citenamefont {Steed}(2016)}]{Jones2016}%
  \BibitemOpen
  \bibfield  {author} {\bibinfo {author} {\bibfnamefont {C.~D.}\ \bibnamefont {Jones}}\ and\ \bibinfo {author} {\bibfnamefont {J.~W.}\ \bibnamefont {Steed}},\ }\href {\doibase 10.1039/c6cs00435k} {\enquote {\bibinfo {title} {{Gels with sense: Supramolecular materials that respond to heat, light and sound}},}\ } (\bibinfo {year} {2016})\BibitemShut {NoStop}%
\bibitem [{\citenamefont {Jiang}\ \emph {et~al.}(2022)\citenamefont {Jiang}, \citenamefont {Makino}, \citenamefont {Royer},\ and\ \citenamefont {Poon}}]{Jiang2022}%
  \BibitemOpen
  \bibfield  {author} {\bibinfo {author} {\bibfnamefont {Y.}~\bibnamefont {Jiang}}, \bibinfo {author} {\bibfnamefont {S.}~\bibnamefont {Makino}}, \bibinfo {author} {\bibfnamefont {J.~R.}\ \bibnamefont {Royer}}, \ and\ \bibinfo {author} {\bibfnamefont {W.~C.}\ \bibnamefont {Poon}},\ }\bibfield  {title} {\enquote {\bibinfo {title} {{Flow-Switched Bistability in a Colloidal Gel with Non-Brownian Grains}},}\ }\href {\doibase 10.1103/PhysRevLett.128.248002} {\bibfield  {journal} {\bibinfo  {journal} {Phys. Rev. Lett.}\ }\textbf {\bibinfo {volume} {128}} (\bibinfo {year} {2022}),\ 10.1103/PhysRevLett.128.248002},\ \Eprint {http://arxiv.org/abs/2111.06019} {2111.06019} \BibitemShut {NoStop}%
\bibitem [{\citenamefont {Nelson}\ \emph {et~al.}(2022)\citenamefont {Nelson}, \citenamefont {Wang}, \citenamefont {Wang}, \citenamefont {Margotta}, \citenamefont {Sammler}, \citenamefont {Izmitli}, \citenamefont {Katz}, \citenamefont {Curtis-Fisk}, \citenamefont {Li},\ and\ \citenamefont {Ewoldt}}]{Nelson2022}%
  \BibitemOpen
  \bibfield  {author} {\bibinfo {author} {\bibfnamefont {A.~Z.}\ \bibnamefont {Nelson}}, \bibinfo {author} {\bibfnamefont {Y.}~\bibnamefont {Wang}}, \bibinfo {author} {\bibfnamefont {Y.}~\bibnamefont {Wang}}, \bibinfo {author} {\bibfnamefont {A.~S.}\ \bibnamefont {Margotta}}, \bibinfo {author} {\bibfnamefont {R.~L.}\ \bibnamefont {Sammler}}, \bibinfo {author} {\bibfnamefont {A.}~\bibnamefont {Izmitli}}, \bibinfo {author} {\bibfnamefont {J.~S.}\ \bibnamefont {Katz}}, \bibinfo {author} {\bibfnamefont {J.}~\bibnamefont {Curtis-Fisk}}, \bibinfo {author} {\bibfnamefont {Y.}~\bibnamefont {Li}}, \ and\ \bibinfo {author} {\bibfnamefont {R.~H.}\ \bibnamefont {Ewoldt}},\ }\bibfield  {title} {\enquote {\bibinfo {title} {Gelation under stress: impact of shear flow on the formation and mechanical properties of methylcellulose hydrogels},}\ }\href {\doibase 10.1039/d1sm01711j} {\bibfield  {journal} {\bibinfo  {journal} {Soft Matter}\ }\textbf {\bibinfo {volume} {18}},\ \bibinfo {pages} {1554--1565} (\bibinfo {year}
  {2022})}\BibitemShut {NoStop}%
\bibitem [{\citenamefont {Koumakis}\ \emph {et~al.}(2015)\citenamefont {Koumakis}, \citenamefont {Moghimi}, \citenamefont {Besseling}, \citenamefont {Poon}, \citenamefont {Brady},\ and\ \citenamefont {Petekidis}}]{Koumakis2015}%
  \BibitemOpen
  \bibfield  {author} {\bibinfo {author} {\bibfnamefont {N.}~\bibnamefont {Koumakis}}, \bibinfo {author} {\bibfnamefont {E.}~\bibnamefont {Moghimi}}, \bibinfo {author} {\bibfnamefont {R.}~\bibnamefont {Besseling}}, \bibinfo {author} {\bibfnamefont {W.~C.}\ \bibnamefont {Poon}}, \bibinfo {author} {\bibfnamefont {J.~F.}\ \bibnamefont {Brady}}, \ and\ \bibinfo {author} {\bibfnamefont {G.}~\bibnamefont {Petekidis}},\ }\bibfield  {title} {\enquote {\bibinfo {title} {{Tuning colloidal gels by shear}},}\ }\href {\doibase 10.1039/c5sm00411j} {\bibfield  {journal} {\bibinfo  {journal} {Soft Matter}\ }\textbf {\bibinfo {volume} {11}},\ \bibinfo {pages} {4640--4648} (\bibinfo {year} {2015})}\BibitemShut {NoStop}%
\bibitem [{\citenamefont {Dag{\`{e}}s}\ \emph {et~al.}(2021)\citenamefont {Dag{\`{e}}s}, \citenamefont {Lidon}, \citenamefont {Jung}, \citenamefont {Pignon}, \citenamefont {Manneville},\ and\ \citenamefont {Gibaud}}]{Dages2021}%
  \BibitemOpen
  \bibfield  {author} {\bibinfo {author} {\bibfnamefont {N.}~\bibnamefont {Dag{\`{e}}s}}, \bibinfo {author} {\bibfnamefont {P.}~\bibnamefont {Lidon}}, \bibinfo {author} {\bibfnamefont {G.}~\bibnamefont {Jung}}, \bibinfo {author} {\bibfnamefont {F.}~\bibnamefont {Pignon}}, \bibinfo {author} {\bibfnamefont {S.}~\bibnamefont {Manneville}}, \ and\ \bibinfo {author} {\bibfnamefont {T.}~\bibnamefont {Gibaud}},\ }\bibfield  {title} {\enquote {\bibinfo {title} {{Mechanics and structure of carbon black gels under high-power ultrasound}},}\ }\href {\doibase 10.1122/8.0000187} {\bibfield  {journal} {\bibinfo  {journal} {J. Rheol. (N. Y. N. Y).}\ }\textbf {\bibinfo {volume} {65}},\ \bibinfo {pages} {477--490} (\bibinfo {year} {2021})},\ \Eprint {http://arxiv.org/abs/2011.06809} {arXiv:2011.06809} \BibitemShut {NoStop}%
\bibitem [{\citenamefont {Tasoglu}\ \emph {et~al.}(2014)\citenamefont {Tasoglu}, \citenamefont {Yu}, \citenamefont {Gungordu}, \citenamefont {Guven}, \citenamefont {Vural},\ and\ \citenamefont {Demirci}}]{tasoglu2014}%
  \BibitemOpen
  \bibfield  {author} {\bibinfo {author} {\bibfnamefont {S.}~\bibnamefont {Tasoglu}}, \bibinfo {author} {\bibfnamefont {C.}~\bibnamefont {Yu}}, \bibinfo {author} {\bibfnamefont {H.}~\bibnamefont {Gungordu}}, \bibinfo {author} {\bibfnamefont {S.}~\bibnamefont {Guven}}, \bibinfo {author} {\bibfnamefont {T.}~\bibnamefont {Vural}}, \ and\ \bibinfo {author} {\bibfnamefont {U.}~\bibnamefont {Demirci}},\ }\bibfield  {title} {\enquote {\bibinfo {title} {Guided and magnetic self-assembly of tunable magnetoceptive gels},}\ }\href@noop {} {\bibfield  {journal} {\bibinfo  {journal} {Nature communications}\ }\textbf {\bibinfo {volume} {5}},\ \bibinfo {pages} {4702} (\bibinfo {year} {2014})}\BibitemShut {NoStop}%
\bibitem [{\citenamefont {Chowdhury}\ \emph {et~al.}(2021)\citenamefont {Chowdhury}, \citenamefont {Reynard-Feytis}, \citenamefont {Roizard}, \citenamefont {Frath}, \citenamefont {Chevallier}, \citenamefont {Bucher},\ and\ \citenamefont {Gibaud}}]{Chowdhury2021}%
  \BibitemOpen
  \bibfield  {author} {\bibinfo {author} {\bibfnamefont {S.}~\bibnamefont {Chowdhury}}, \bibinfo {author} {\bibfnamefont {Q.}~\bibnamefont {Reynard-Feytis}}, \bibinfo {author} {\bibfnamefont {C.}~\bibnamefont {Roizard}}, \bibinfo {author} {\bibfnamefont {D.}~\bibnamefont {Frath}}, \bibinfo {author} {\bibfnamefont {F.}~\bibnamefont {Chevallier}}, \bibinfo {author} {\bibfnamefont {C.}~\bibnamefont {Bucher}}, \ and\ \bibinfo {author} {\bibfnamefont {T.}~\bibnamefont {Gibaud}},\ }\bibfield  {title} {\enquote {\bibinfo {title} {{Light-Controlled Aggregation and Gelation of Viologen-Based Coordination Polymers}},}\ }\href {\doibase 10.1021/acs.jpcb.1c06090} {\bibfield  {journal} {\bibinfo  {journal} {J. Phys. Chem. B}\ }\textbf {\bibinfo {volume} {125}},\ \bibinfo {pages} {12063--12071} (\bibinfo {year} {2021})},\ \Eprint {http://arxiv.org/abs/2106.15688} {2106.15688} \BibitemShut {NoStop}%
\bibitem [{\citenamefont {Fiocco}, \citenamefont {Foffi},\ and\ \citenamefont {Sastry}(2014)}]{Fiocco2014}%
  \BibitemOpen
  \bibfield  {author} {\bibinfo {author} {\bibfnamefont {D.}~\bibnamefont {Fiocco}}, \bibinfo {author} {\bibfnamefont {G.}~\bibnamefont {Foffi}}, \ and\ \bibinfo {author} {\bibfnamefont {S.}~\bibnamefont {Sastry}},\ }\bibfield  {title} {\enquote {\bibinfo {title} {Encoding of memory in sheared amorphous solids},}\ }\href {\doibase 10.1103/PHYSREVLETT.112.025702/FIGURES/4/MEDIUM} {\bibfield  {journal} {\bibinfo  {journal} {Physical Review Letters}\ }\textbf {\bibinfo {volume} {112}},\ \bibinfo {pages} {025702} (\bibinfo {year} {2014})}\BibitemShut {NoStop}%
\bibitem [{\citenamefont {Mewis}\ and\ \citenamefont {Wagner}(2009)}]{Mewis2009}%
  \BibitemOpen
  \bibfield  {author} {\bibinfo {author} {\bibfnamefont {J.}~\bibnamefont {Mewis}}\ and\ \bibinfo {author} {\bibfnamefont {N.~J.}\ \bibnamefont {Wagner}},\ }\bibfield  {title} {\enquote {\bibinfo {title} {{Thixotropy}},}\ }\href {\doibase 10.1016/j.cis.2008.09.005} {\bibfield  {journal} {\bibinfo  {journal} {Adv. Colloid Interface Sci.}\ }\textbf {\bibinfo {volume} {147-148}},\ \bibinfo {pages} {214--227} (\bibinfo {year} {2009})}\BibitemShut {NoStop}%
\bibitem [{\citenamefont {Varga}\ and\ \citenamefont {Swan}(2018)}]{Varga2018}%
  \BibitemOpen
  \bibfield  {author} {\bibinfo {author} {\bibfnamefont {Z.}~\bibnamefont {Varga}}\ and\ \bibinfo {author} {\bibfnamefont {J.~W.}\ \bibnamefont {Swan}},\ }\bibfield  {title} {\enquote {\bibinfo {title} {{Large scale anisotropies in sheared colloidal gels}},}\ }\href {\doibase 10.1122/1.5003364} {\bibfield  {journal} {\bibinfo  {journal} {J. Rheol. (N. Y. N. Y).}\ }\textbf {\bibinfo {volume} {62}},\ \bibinfo {pages} {405--418} (\bibinfo {year} {2018})}\BibitemShut {NoStop}%
\bibitem [{\citenamefont {Jamali}, \citenamefont {Armstrong},\ and\ \citenamefont {McKinley}(2020)}]{Jamali2020}%
  \BibitemOpen
  \bibfield  {author} {\bibinfo {author} {\bibfnamefont {S.}~\bibnamefont {Jamali}}, \bibinfo {author} {\bibfnamefont {R.~C.}\ \bibnamefont {Armstrong}}, \ and\ \bibinfo {author} {\bibfnamefont {G.~H.}\ \bibnamefont {McKinley}},\ }\bibfield  {title} {\enquote {\bibinfo {title} {{Time-rate-transformation framework for targeted assembly of short-range attractive colloidal suspensions}},}\ }\href {\doibase 10.1016/j.mtadv.2019.100026} {\bibfield  {journal} {\bibinfo  {journal} {Mater. Today Adv.}\ }\textbf {\bibinfo {volume} {5}},\ \bibinfo {pages} {100026} (\bibinfo {year} {2020})}\BibitemShut {NoStop}%
\bibitem [{\citenamefont {Moghimi}\ \emph {et~al.}(2017)\citenamefont {Moghimi}, \citenamefont {Jacob}, \citenamefont {Koumakis},\ and\ \citenamefont {Petekidis}}]{Moghimi2017}%
  \BibitemOpen
  \bibfield  {author} {\bibinfo {author} {\bibfnamefont {E.}~\bibnamefont {Moghimi}}, \bibinfo {author} {\bibfnamefont {A.~R.}\ \bibnamefont {Jacob}}, \bibinfo {author} {\bibfnamefont {N.}~\bibnamefont {Koumakis}}, \ and\ \bibinfo {author} {\bibfnamefont {G.}~\bibnamefont {Petekidis}},\ }\bibfield  {title} {\enquote {\bibinfo {title} {{Colloidal gels tuned by oscillatory shear}},}\ }\href {\doibase 10.1039/c6sm02508k} {\bibfield  {journal} {\bibinfo  {journal} {Soft Matter}\ }\textbf {\bibinfo {volume} {13}},\ \bibinfo {pages} {2371--2383} (\bibinfo {year} {2017})}\BibitemShut {NoStop}%
\bibitem [{\citenamefont {Das}\ and\ \citenamefont {Petekidis}(2022)}]{Das2022}%
  \BibitemOpen
  \bibfield  {author} {\bibinfo {author} {\bibfnamefont {M.}~\bibnamefont {Das}}\ and\ \bibinfo {author} {\bibfnamefont {G.}~\bibnamefont {Petekidis}},\ }\bibfield  {title} {\enquote {\bibinfo {title} {{Shear induced tuning and memory effects in colloidal gels of rods and spheres}},}\ }\href {\doibase 10.1063/5.0129709} {\bibfield  {journal} {\bibinfo  {journal} {J. Chem. Phys.}\ }\textbf {\bibinfo {volume} {157}} (\bibinfo {year} {2022}),\ 10.1063/5.0129709}\BibitemShut {NoStop}%
\bibitem [{\citenamefont {Rocklin}\ \emph {et~al.}(2021)\citenamefont {Rocklin}, \citenamefont {Hsiao}, \citenamefont {Szakasits}, \citenamefont {Solomon},\ and\ \citenamefont {Mao}}]{Rocklin2021}%
  \BibitemOpen
  \bibfield  {author} {\bibinfo {author} {\bibfnamefont {D.~Z.}\ \bibnamefont {Rocklin}}, \bibinfo {author} {\bibfnamefont {L.}~\bibnamefont {Hsiao}}, \bibinfo {author} {\bibfnamefont {M.}~\bibnamefont {Szakasits}}, \bibinfo {author} {\bibfnamefont {M.~J.}\ \bibnamefont {Solomon}}, \ and\ \bibinfo {author} {\bibfnamefont {X.}~\bibnamefont {Mao}},\ }\bibfield  {title} {\enquote {\bibinfo {title} {{Elasticity of colloidal gels: structural heterogeneity, floppy modes, and rigidity}},}\ }\href {\doibase 10.1039/d0sm00053a} {\bibfield  {journal} {\bibinfo  {journal} {Soft Matter}\ }\textbf {\bibinfo {volume} {17}},\ \bibinfo {pages} {6929--6934} (\bibinfo {year} {2021})},\ \Eprint {http://arxiv.org/abs/1808.01533} {1808.01533} \BibitemShut {NoStop}%
\bibitem [{\citenamefont {Sudreau}\ \emph {et~al.}(2022)\citenamefont {Sudreau}, \citenamefont {Auxois}, \citenamefont {Servel}, \citenamefont {L{\'{e}}colier}, \citenamefont {Manneville},\ and\ \citenamefont {Divoux}}]{Sudreau2022}%
  \BibitemOpen
  \bibfield  {author} {\bibinfo {author} {\bibfnamefont {I.}~\bibnamefont {Sudreau}}, \bibinfo {author} {\bibfnamefont {M.}~\bibnamefont {Auxois}}, \bibinfo {author} {\bibfnamefont {M.}~\bibnamefont {Servel}}, \bibinfo {author} {\bibfnamefont {{\'{E}}.}~\bibnamefont {L{\'{e}}colier}}, \bibinfo {author} {\bibfnamefont {S.}~\bibnamefont {Manneville}}, \ and\ \bibinfo {author} {\bibfnamefont {T.}~\bibnamefont {Divoux}},\ }\bibfield  {title} {\enquote {\bibinfo {title} {{Residual stresses and shear-induced overaging in boehmite gels}},}\ }\href {\doibase 10.1103/PhysRevMaterials.6.L042601} {\bibfield  {journal} {\bibinfo  {journal} {Phys. Rev. Mater.}\ }\textbf {\bibinfo {volume} {6}} (\bibinfo {year} {2022}),\ 10.1103/PhysRevMaterials.6.L042601},\ \Eprint {http://arxiv.org/abs/2201.02528} {2201.02528} \BibitemShut {NoStop}%
\bibitem [{\citenamefont {Dag{\`{e}}s}\ \emph {et~al.}(2022)\citenamefont {Dag{\`{e}}s}, \citenamefont {Bouthier}, \citenamefont {Matthews}, \citenamefont {Manneville}, \citenamefont {Divoux}, \citenamefont {Poulesquen},\ and\ \citenamefont {Gibaud}}]{Dages2022b}%
  \BibitemOpen
  \bibfield  {author} {\bibinfo {author} {\bibfnamefont {N.}~\bibnamefont {Dag{\`{e}}s}}, \bibinfo {author} {\bibfnamefont {L.~V.}\ \bibnamefont {Bouthier}}, \bibinfo {author} {\bibfnamefont {L.}~\bibnamefont {Matthews}}, \bibinfo {author} {\bibfnamefont {S.}~\bibnamefont {Manneville}}, \bibinfo {author} {\bibfnamefont {T.}~\bibnamefont {Divoux}}, \bibinfo {author} {\bibfnamefont {A.}~\bibnamefont {Poulesquen}}, \ and\ \bibinfo {author} {\bibfnamefont {T.}~\bibnamefont {Gibaud}},\ }\bibfield  {title} {\enquote {\bibinfo {title} {{Interpenetration of fractal clusters drives elasticity in colloidal gels formed upon flow cessation}},}\ }\href {\doibase 10.1039/d2sm00481j} {\bibfield  {journal} {\bibinfo  {journal} {Soft Matter}\ }\textbf {\bibinfo {volume} {18}},\ \bibinfo {pages} {6645--6659} (\bibinfo {year} {2022})}\BibitemShut {NoStop}%
\bibitem [{\citenamefont {Colombo}\ \emph {et~al.}(2017)\citenamefont {Colombo}, \citenamefont {Kim}, \citenamefont {Schweizer}, \citenamefont {Schroyen}, \citenamefont {Clasen}, \citenamefont {Mewis},\ and\ \citenamefont {Vermant}}]{Colombo2017}%
  \BibitemOpen
  \bibfield  {author} {\bibinfo {author} {\bibfnamefont {G.}~\bibnamefont {Colombo}}, \bibinfo {author} {\bibfnamefont {S.}~\bibnamefont {Kim}}, \bibinfo {author} {\bibfnamefont {T.}~\bibnamefont {Schweizer}}, \bibinfo {author} {\bibfnamefont {B.}~\bibnamefont {Schroyen}}, \bibinfo {author} {\bibfnamefont {C.}~\bibnamefont {Clasen}}, \bibinfo {author} {\bibfnamefont {J.}~\bibnamefont {Mewis}}, \ and\ \bibinfo {author} {\bibfnamefont {J.}~\bibnamefont {Vermant}},\ }\bibfield  {title} {\enquote {\bibinfo {title} {{Superposition rheology and anisotropy in rheological properties of sheared colloidal gels}},}\ }\href {\doibase 10.1122/1.4998176} {\bibfield  {journal} {\bibinfo  {journal} {J. Rheol. (N. Y. N. Y).}\ }\textbf {\bibinfo {volume} {61}},\ \bibinfo {pages} {1035--1048} (\bibinfo {year} {2017})}\BibitemShut {NoStop}%
\bibitem [{\citenamefont {Hsiao}\ \emph {et~al.}(2012)\citenamefont {Hsiao}, \citenamefont {Newman}, \citenamefont {Glotzer},\ and\ \citenamefont {Solomon}}]{Hsiao2012}%
  \BibitemOpen
  \bibfield  {author} {\bibinfo {author} {\bibfnamefont {L.~C.}\ \bibnamefont {Hsiao}}, \bibinfo {author} {\bibfnamefont {R.~S.}\ \bibnamefont {Newman}}, \bibinfo {author} {\bibfnamefont {S.~C.}\ \bibnamefont {Glotzer}}, \ and\ \bibinfo {author} {\bibfnamefont {M.~J.}\ \bibnamefont {Solomon}},\ }\bibfield  {title} {\enquote {\bibinfo {title} {{Role of isostaticity and load-bearing microstructure in the elasticity of yielded colloidal gels}},}\ }\href {\doibase 10.1073/pnas.1206742109} {\bibfield  {journal} {\bibinfo  {journal} {Proc. Natl. Acad. Sci. U. S. A.}\ }\textbf {\bibinfo {volume} {109}},\ \bibinfo {pages} {16029--16034} (\bibinfo {year} {2012})}\BibitemShut {NoStop}%
\bibitem [{\citenamefont {Shih}\ \emph {et~al.}(1990)\citenamefont {Shih}, \citenamefont {Shih}, \citenamefont {Kim}, \citenamefont {Liu},\ and\ \citenamefont {Aksay}}]{Shih1990}%
  \BibitemOpen
  \bibfield  {author} {\bibinfo {author} {\bibfnamefont {W.~H. W.~Y.}\ \bibnamefont {Shih}}, \bibinfo {author} {\bibfnamefont {W.~H. W.~Y.}\ \bibnamefont {Shih}}, \bibinfo {author} {\bibfnamefont {S.~I.}\ \bibnamefont {Kim}}, \bibinfo {author} {\bibfnamefont {J.}~\bibnamefont {Liu}}, \ and\ \bibinfo {author} {\bibfnamefont {I.~A.}\ \bibnamefont {Aksay}},\ }\bibfield  {title} {\enquote {\bibinfo {title} {{Scaling behavior of the elastic properties of colloidal gels}},}\ }\href {\doibase 10.1103/PhysRevA.42.4772} {\bibfield  {journal} {\bibinfo  {journal} {Phys. Rev. A}\ }\textbf {\bibinfo {volume} {42}},\ \bibinfo {pages} {4772--4779} (\bibinfo {year} {1990})}\BibitemShut {NoStop}%
\bibitem [{\citenamefont {Wu}\ and\ \citenamefont {Morbidelli}(2001)}]{Wu2001}%
  \BibitemOpen
  \bibfield  {author} {\bibinfo {author} {\bibfnamefont {H.}~\bibnamefont {Wu}}\ and\ \bibinfo {author} {\bibfnamefont {M.}~\bibnamefont {Morbidelli}},\ }\bibfield  {title} {\enquote {\bibinfo {title} {{Model relating structure of colloidal gels to their elastic properties}},}\ }\href {\doibase 10.1021/la001121f} {\bibfield  {journal} {\bibinfo  {journal} {Langmuir}\ }\textbf {\bibinfo {volume} {17}},\ \bibinfo {pages} {1030--1036} (\bibinfo {year} {2001})}\BibitemShut {NoStop}%
\bibitem [{\citenamefont {Bouthier}\ and\ \citenamefont {Gibaud}(2023)}]{Bouthier2022b}%
  \BibitemOpen
  \bibfield  {author} {\bibinfo {author} {\bibfnamefont {L.-V.}\ \bibnamefont {Bouthier}}\ and\ \bibinfo {author} {\bibfnamefont {T.}~\bibnamefont {Gibaud}},\ }\bibfield  {title} {\enquote {\bibinfo {title} {{Three length scales colloidal gels: the clusters of clusters versus the interpenetrating clusters approach}},}\ }\href {\doibase 10.1122/8.0000595} {\bibfield  {journal} {\bibinfo  {journal} {J. Rheol. (N. Y. N. Y).}\ }\textbf {\bibinfo {volume} {67}},\ \bibinfo {pages} {621--633} (\bibinfo {year} {2023})}\BibitemShut {NoStop}%
\bibitem [{\citenamefont {Zaccone}\ \emph {et~al.}(2009)\citenamefont {Zaccone}, \citenamefont {Soos}, \citenamefont {Lattuada}, \citenamefont {Wu}, \citenamefont {B{\"{a}}bler},\ and\ \citenamefont {Morbidelli}}]{Zaccone2009}%
  \BibitemOpen
  \bibfield  {author} {\bibinfo {author} {\bibfnamefont {A.}~\bibnamefont {Zaccone}}, \bibinfo {author} {\bibfnamefont {M.}~\bibnamefont {Soos}}, \bibinfo {author} {\bibfnamefont {M.}~\bibnamefont {Lattuada}}, \bibinfo {author} {\bibfnamefont {H.}~\bibnamefont {Wu}}, \bibinfo {author} {\bibfnamefont {M.~U.}\ \bibnamefont {B{\"{a}}bler}}, \ and\ \bibinfo {author} {\bibfnamefont {M.}~\bibnamefont {Morbidelli}},\ }\bibfield  {title} {\enquote {\bibinfo {title} {{Breakup of dense colloidal aggregates under hydrodynamic stresses}},}\ }\href {\doibase 10.1103/PhysRevE.79.061401} {\bibfield  {journal} {\bibinfo  {journal} {Phys. Rev. E - Stat. Nonlinear, Soft Matter Phys.}\ }\textbf {\bibinfo {volume} {79}},\ \bibinfo {pages} {061401} (\bibinfo {year} {2009})}\BibitemShut {NoStop}%
\bibitem [{\citenamefont {Whitaker}\ \emph {et~al.}(2019)\citenamefont {Whitaker}, \citenamefont {Varga}, \citenamefont {Hsiao}, \citenamefont {Solomon}, \citenamefont {Swan},\ and\ \citenamefont {Furst}}]{Whitaker2019}%
  \BibitemOpen
  \bibfield  {author} {\bibinfo {author} {\bibfnamefont {K.~A.}\ \bibnamefont {Whitaker}}, \bibinfo {author} {\bibfnamefont {Z.}~\bibnamefont {Varga}}, \bibinfo {author} {\bibfnamefont {L.~C.}\ \bibnamefont {Hsiao}}, \bibinfo {author} {\bibfnamefont {M.~J.}\ \bibnamefont {Solomon}}, \bibinfo {author} {\bibfnamefont {J.~W.}\ \bibnamefont {Swan}}, \ and\ \bibinfo {author} {\bibfnamefont {E.~M.}\ \bibnamefont {Furst}},\ }\bibfield  {title} {\enquote {\bibinfo {title} {{Colloidal gel elasticity arises from the packing of locally glassy clusters}},}\ }\href {\doibase 10.1038/s41467-019-10039-w} {\bibfield  {journal} {\bibinfo  {journal} {Nat. Commun.}\ }\textbf {\bibinfo {volume} {10}} (\bibinfo {year} {2019}),\ 10.1038/s41467-019-10039-w}\BibitemShut {NoStop}%
\bibitem [{\citenamefont {Nabizadeh}\ \emph {et~al.}(2024)\citenamefont {Nabizadeh}, \citenamefont {Nasirian}, \citenamefont {Li}, \citenamefont {Saraswat}, \citenamefont {Waheibi}, \citenamefont {Hsiao}, \citenamefont {Bi}, \citenamefont {Ravandi},\ and\ \citenamefont {Jamali}}]{Nabizadeh2024}%
  \BibitemOpen
  \bibfield  {author} {\bibinfo {author} {\bibfnamefont {M.}~\bibnamefont {Nabizadeh}}, \bibinfo {author} {\bibfnamefont {F.}~\bibnamefont {Nasirian}}, \bibinfo {author} {\bibfnamefont {X.}~\bibnamefont {Li}}, \bibinfo {author} {\bibfnamefont {Y.}~\bibnamefont {Saraswat}}, \bibinfo {author} {\bibfnamefont {R.}~\bibnamefont {Waheibi}}, \bibinfo {author} {\bibfnamefont {L.~C.}\ \bibnamefont {Hsiao}}, \bibinfo {author} {\bibfnamefont {D.}~\bibnamefont {Bi}}, \bibinfo {author} {\bibfnamefont {B.}~\bibnamefont {Ravandi}}, \ and\ \bibinfo {author} {\bibfnamefont {S.}~\bibnamefont {Jamali}},\ }\bibfield  {title} {\enquote {\bibinfo {title} {Network physics of attractive colloidal gels: Resilience, rigidity, and phase diagram},}\ }\href {\doibase 10.1073/pnas.2316394121} {\bibfield  {journal} {\bibinfo  {journal} {Proceedings of the National Academy of Sciences of the United States of America}\ }\textbf {\bibinfo {volume} {121}},\ \bibinfo {pages} {1--11} (\bibinfo {year} {2024})}\BibitemShut {NoStop}%
\bibitem [{\citenamefont {Bauland}\ \emph {et~al.}(2024)\citenamefont {Bauland}, \citenamefont {Bouthier}, \citenamefont {Poulesquen},\ and\ \citenamefont {Gibaud}}]{Bauland2024}%
  \BibitemOpen
  \bibfield  {author} {\bibinfo {author} {\bibfnamefont {J.}~\bibnamefont {Bauland}}, \bibinfo {author} {\bibfnamefont {L.-V.}\ \bibnamefont {Bouthier}}, \bibinfo {author} {\bibfnamefont {A.}~\bibnamefont {Poulesquen}}, \ and\ \bibinfo {author} {\bibfnamefont {T.}~\bibnamefont {Gibaud}},\ }\bibfield  {title} {\enquote {\bibinfo {title} {{Attractive carbon black dispersions: Structural and mechanical responses to shear}},}\ }\href {\doibase 10.1122/8.0000791} {\bibfield  {journal} {\bibinfo  {journal} {J. Rheol}\ }\textbf {\bibinfo {volume} {68}},\ \bibinfo {pages} {429--443} (\bibinfo {year} {2024})}\BibitemShut {NoStop}%
\bibitem [{\citenamefont {Bauland}\ \emph {et~al.}(2025)\citenamefont {Bauland}, \citenamefont {Legrand}, \citenamefont {Manneville}, \citenamefont {Divoux}, \citenamefont {Poulesquen},\ and\ \citenamefont {Gibaud}}]{Bauland2025}%
  \BibitemOpen
  \bibfield  {author} {\bibinfo {author} {\bibfnamefont {J.}~\bibnamefont {Bauland}}, \bibinfo {author} {\bibfnamefont {G.}~\bibnamefont {Legrand}}, \bibinfo {author} {\bibfnamefont {S.}~\bibnamefont {Manneville}}, \bibinfo {author} {\bibfnamefont {T.}~\bibnamefont {Divoux}}, \bibinfo {author} {\bibfnamefont {A.}~\bibnamefont {Poulesquen}}, \ and\ \bibinfo {author} {\bibfnamefont {T.}~\bibnamefont {Gibaud}},\ }\bibfield  {title} {\enquote {\bibinfo {title} {Antithixotropic dynamics in attractive colloidal dispersions: A shear restructuring driven by elastic stresses},}\ }\href {\doibase 10.1122/8.0000980} {\bibfield  {journal} {\bibinfo  {journal} {Journal of Rheology}\ }\textbf {\bibinfo {volume} {69}},\ \bibinfo {pages} {583--598} (\bibinfo {year} {2025})}\BibitemShut {NoStop}%
\bibitem [{\citenamefont {Narayanan}\ \emph {et~al.}(2020)\citenamefont {Narayanan}, \citenamefont {Dattani}, \citenamefont {M{\"o}ller},\ and\ \citenamefont {Kwa{\'s}niewski}}]{narayanan2020}%
  \BibitemOpen
  \bibfield  {author} {\bibinfo {author} {\bibfnamefont {T.}~\bibnamefont {Narayanan}}, \bibinfo {author} {\bibfnamefont {R.}~\bibnamefont {Dattani}}, \bibinfo {author} {\bibfnamefont {J.}~\bibnamefont {M{\"o}ller}}, \ and\ \bibinfo {author} {\bibfnamefont {P.}~\bibnamefont {Kwa{\'s}niewski}},\ }\bibfield  {title} {\enquote {\bibinfo {title} {A microvolume shear cell for combined rheology and x-ray scattering experiments},}\ }\href@noop {} {\bibfield  {journal} {\bibinfo  {journal} {Review of Scientific Instruments}\ }\textbf {\bibinfo {volume} {91}} (\bibinfo {year} {2020})}\BibitemShut {NoStop}%
\bibitem [{\citenamefont {Hipp}, \citenamefont {Richards},\ and\ \citenamefont {Wagner}(2019)}]{Hipp2019}%
  \BibitemOpen
  \bibfield  {author} {\bibinfo {author} {\bibfnamefont {J.~B.}\ \bibnamefont {Hipp}}, \bibinfo {author} {\bibfnamefont {J.~J.}\ \bibnamefont {Richards}}, \ and\ \bibinfo {author} {\bibfnamefont {N.~J.}\ \bibnamefont {Wagner}},\ }\bibfield  {title} {\enquote {\bibinfo {title} {{Structure-property relationships of sheared carbon black suspensions determined by simultaneous rheological and neutron scattering measurements}},}\ }\href {\doibase 10.1122/1.5071470} {\bibfield  {journal} {\bibinfo  {journal} {J. Rheol. (N. Y. N. Y).}\ }\textbf {\bibinfo {volume} {63}},\ \bibinfo {pages} {423--436} (\bibinfo {year} {2019})}\BibitemShut {NoStop}%
\bibitem [{\citenamefont {Wang}\ and\ \citenamefont {Ewoldt}(2022{\natexlab{a}})}]{Wang2022}%
  \BibitemOpen
  \bibfield  {author} {\bibinfo {author} {\bibfnamefont {Y.}~\bibnamefont {Wang}}\ and\ \bibinfo {author} {\bibfnamefont {R.~H.}\ \bibnamefont {Ewoldt}},\ }\bibfield  {title} {\enquote {\bibinfo {title} {{New insights on carbon black suspension rheology -- anisotropic thixotropy and anti-thixotropy}},}\ }\href@noop {} {\bibfield  {journal} {\bibinfo  {journal} {J. Rheol. (N. Y. N. Y).}\ }\textbf {\bibinfo {volume} {66}},\ \bibinfo {pages} {937--953} (\bibinfo {year} {2022}{\natexlab{a}})},\ \Eprint {http://arxiv.org/abs/2202.05772} {2202.05772} \BibitemShut {NoStop}%
\bibitem [{\citenamefont {Trappe}\ and\ \citenamefont {Weitz}(2000)}]{Trappe2000}%
  \BibitemOpen
  \bibfield  {author} {\bibinfo {author} {\bibfnamefont {V.}~\bibnamefont {Trappe}}\ and\ \bibinfo {author} {\bibfnamefont {D.~A.}\ \bibnamefont {Weitz}},\ }\bibfield  {title} {\enquote {\bibinfo {title} {{Scaling of the viscoelasticity of weakly attractive particles}},}\ }\href {\doibase 10.1103/PhysRevLett.85.449} {\bibfield  {journal} {\bibinfo  {journal} {Phys. Rev. Lett.}\ }\textbf {\bibinfo {volume} {85}},\ \bibinfo {pages} {449--452} (\bibinfo {year} {2000})}\BibitemShut {NoStop}%
\bibitem [{\citenamefont {Hipp}, \citenamefont {Richards},\ and\ \citenamefont {Wagner}(2021)}]{Hipp2021}%
  \BibitemOpen
  \bibfield  {author} {\bibinfo {author} {\bibfnamefont {J.~B.}\ \bibnamefont {Hipp}}, \bibinfo {author} {\bibfnamefont {J.~J.}\ \bibnamefont {Richards}}, \ and\ \bibinfo {author} {\bibfnamefont {N.~J.}\ \bibnamefont {Wagner}},\ }\bibfield  {title} {\enquote {\bibinfo {title} {{Direct measurements of the microstructural origin of shear-thinning in carbon black suspensions}},}\ }\href {\doibase 10.1122/8.0000089} {\bibfield  {journal} {\bibinfo  {journal} {J. Rheol. (N. Y. N. Y).}\ }\textbf {\bibinfo {volume} {65}},\ \bibinfo {pages} {145--157} (\bibinfo {year} {2021})}\BibitemShut {NoStop}%
\bibitem [{\citenamefont {Bouthier}\ \emph {et~al.}(2023)\citenamefont {Bouthier}, \citenamefont {Castellani}, \citenamefont {Manneville}, \citenamefont {Poulesquen}, \citenamefont {Valette},\ and\ \citenamefont {Hachem}}]{Bouthier2023}%
  \BibitemOpen
  \bibfield  {author} {\bibinfo {author} {\bibfnamefont {L.~V.}\ \bibnamefont {Bouthier}}, \bibinfo {author} {\bibfnamefont {R.}~\bibnamefont {Castellani}}, \bibinfo {author} {\bibfnamefont {S.}~\bibnamefont {Manneville}}, \bibinfo {author} {\bibfnamefont {A.}~\bibnamefont {Poulesquen}}, \bibinfo {author} {\bibfnamefont {R.}~\bibnamefont {Valette}}, \ and\ \bibinfo {author} {\bibfnamefont {E.}~\bibnamefont {Hachem}},\ }\bibfield  {title} {\enquote {\bibinfo {title} {Aggregation and disaggregation processes in clusters of particles under flow: Simple numerical and theoretical insights},}\ }\href {\doibase 10.1103/PHYSREVFLUIDS.8.023304} {\bibfield  {journal} {\bibinfo  {journal} {Physical Review Fluids}\ }\textbf {\bibinfo {volume} {8}} (\bibinfo {year} {2023}),\ 10.1103/PHYSREVFLUIDS.8.023304}\BibitemShut {NoStop}%
\bibitem [{\citenamefont {Lazzari}\ \emph {et~al.}(2016)\citenamefont {Lazzari}, \citenamefont {Nicoud}, \citenamefont {Jaquet}, \citenamefont {Lattuada},\ and\ \citenamefont {Morbidelli}}]{Lazzari2016}%
  \BibitemOpen
  \bibfield  {author} {\bibinfo {author} {\bibfnamefont {S.}~\bibnamefont {Lazzari}}, \bibinfo {author} {\bibfnamefont {L.}~\bibnamefont {Nicoud}}, \bibinfo {author} {\bibfnamefont {B.}~\bibnamefont {Jaquet}}, \bibinfo {author} {\bibfnamefont {M.}~\bibnamefont {Lattuada}}, \ and\ \bibinfo {author} {\bibfnamefont {M.}~\bibnamefont {Morbidelli}},\ }\bibfield  {title} {\enquote {\bibinfo {title} {{Fractal-like structures in colloid science}},}\ }\href {\doibase 10.1016/j.cis.2016.05.002} {\bibfield  {journal} {\bibinfo  {journal} {Adv. Colloid Interface Sci.}\ }\textbf {\bibinfo {volume} {235}},\ \bibinfo {pages} {1--13} (\bibinfo {year} {2016})}\BibitemShut {NoStop}%
\bibitem [{\citenamefont {Dagastine}, \citenamefont {Prieve},\ and\ \citenamefont {White}(2002)}]{dagastine2002}%
  \BibitemOpen
  \bibfield  {author} {\bibinfo {author} {\bibfnamefont {R.~R.}\ \bibnamefont {Dagastine}}, \bibinfo {author} {\bibfnamefont {D.~C.}\ \bibnamefont {Prieve}}, \ and\ \bibinfo {author} {\bibfnamefont {L.~R.}\ \bibnamefont {White}},\ }\bibfield  {title} {\enquote {\bibinfo {title} {Calculations of van der waals forces in 2-dimensionally anisotropic materials and its application to carbon black},}\ }\href@noop {} {\bibfield  {journal} {\bibinfo  {journal} {Journal of colloid and interface science}\ }\textbf {\bibinfo {volume} {249}},\ \bibinfo {pages} {78--83} (\bibinfo {year} {2002})}\BibitemShut {NoStop}%
\bibitem [{\citenamefont {Varga}\ \emph {et~al.}(2019)\citenamefont {Varga}, \citenamefont {Grenard}, \citenamefont {Pecorario}, \citenamefont {Taberlet}, \citenamefont {Dolique}, \citenamefont {Manneville}, \citenamefont {Divoux}, \citenamefont {McKinley},\ and\ \citenamefont {Swan}}]{Varga2019}%
  \BibitemOpen
  \bibfield  {author} {\bibinfo {author} {\bibfnamefont {Z.}~\bibnamefont {Varga}}, \bibinfo {author} {\bibfnamefont {V.}~\bibnamefont {Grenard}}, \bibinfo {author} {\bibfnamefont {S.}~\bibnamefont {Pecorario}}, \bibinfo {author} {\bibfnamefont {N.}~\bibnamefont {Taberlet}}, \bibinfo {author} {\bibfnamefont {V.}~\bibnamefont {Dolique}}, \bibinfo {author} {\bibfnamefont {S.}~\bibnamefont {Manneville}}, \bibinfo {author} {\bibfnamefont {T.}~\bibnamefont {Divoux}}, \bibinfo {author} {\bibfnamefont {G.~H.}\ \bibnamefont {McKinley}}, \ and\ \bibinfo {author} {\bibfnamefont {J.~W.}\ \bibnamefont {Swan}},\ }\bibfield  {title} {\enquote {\bibinfo {title} {{Hydrodynamics control shear-induced pattern formation in attractive suspensions}},}\ }\href {\doibase 10.1073/pnas.1901370116} {\bibfield  {journal} {\bibinfo  {journal} {Proc. Natl. Acad. Sci. U. S. A.}\ }\textbf {\bibinfo {volume} {116}},\ \bibinfo {pages} {12193--12198} (\bibinfo {year} {2019})}\BibitemShut {NoStop}%
\bibitem [{\citenamefont {Trappe}\ \emph {et~al.}(2007)\citenamefont {Trappe}, \citenamefont {Pitard}, \citenamefont {Ramos}, \citenamefont {Robert}, \citenamefont {Bissig},\ and\ \citenamefont {Cipelletti}}]{Trappe2007}%
  \BibitemOpen
  \bibfield  {author} {\bibinfo {author} {\bibfnamefont {V.}~\bibnamefont {Trappe}}, \bibinfo {author} {\bibfnamefont {E.}~\bibnamefont {Pitard}}, \bibinfo {author} {\bibfnamefont {L.}~\bibnamefont {Ramos}}, \bibinfo {author} {\bibfnamefont {A.}~\bibnamefont {Robert}}, \bibinfo {author} {\bibfnamefont {H.}~\bibnamefont {Bissig}}, \ and\ \bibinfo {author} {\bibfnamefont {L.}~\bibnamefont {Cipelletti}},\ }\bibfield  {title} {\enquote {\bibinfo {title} {{Investigation of q -dependent dynamical heterogeneity in a colloidal gel by x-ray photon correlation spectroscopy}},}\ }\href {\doibase 10.1103/PhysRevE.76.051404} {\bibfield  {journal} {\bibinfo  {journal} {Phys. Rev. E - Stat. Nonlinear, Soft Matter Phys.}\ }\textbf {\bibinfo {volume} {76}} (\bibinfo {year} {2007}),\ 10.1103/PhysRevE.76.051404}\BibitemShut {NoStop}%
\bibitem [{\citenamefont {Wang}\ and\ \citenamefont {Ewoldt}(2022{\natexlab{b}})}]{Wang2022a}%
  \BibitemOpen
  \bibfield  {author} {\bibinfo {author} {\bibfnamefont {Y.}~\bibnamefont {Wang}}\ and\ \bibinfo {author} {\bibfnamefont {R.~H.}\ \bibnamefont {Ewoldt}},\ }\bibfield  {title} {\enquote {\bibinfo {title} {{New insights on carbon black suspension rheology—Anisotropic thixotropy and antithixotropy}},}\ }\href {\doibase 10.1122/8.0000455} {\bibfield  {journal} {\bibinfo  {journal} {J. Rheol. (N. Y. N. Y).}\ }\textbf {\bibinfo {volume} {66}},\ \bibinfo {pages} {937--953} (\bibinfo {year} {2022}{\natexlab{b}})}\BibitemShut {NoStop}%
\bibitem [{\citenamefont {Bouthier}\ \emph {et~al.}(2022)\citenamefont {Bouthier}, \citenamefont {Castellani}, \citenamefont {Hachem},\ and\ \citenamefont {Valette}}]{Bouthier2022}%
  \BibitemOpen
  \bibfield  {author} {\bibinfo {author} {\bibfnamefont {L.-V.}\ \bibnamefont {Bouthier}}, \bibinfo {author} {\bibfnamefont {R.}~\bibnamefont {Castellani}}, \bibinfo {author} {\bibfnamefont {E.}~\bibnamefont {Hachem}}, \ and\ \bibinfo {author} {\bibfnamefont {R.}~\bibnamefont {Valette}},\ }\bibfield  {title} {\enquote {\bibinfo {title} {Proposition of extension of models relating rheological quantities and microscopic structure through the use of a double fractal structure},}\ }\href {\doibase 10.1063/5.0101750} {\bibfield  {journal} {\bibinfo  {journal} {Physics of Fluids}\ }\textbf {\bibinfo {volume} {34}},\ \bibinfo {pages} {083105} (\bibinfo {year} {2022})}\BibitemShut {NoStop}%
\bibitem [{\citenamefont {Zaccone}, \citenamefont {Wu},\ and\ \citenamefont {{Del Gado}}(2009)}]{Zaccone2009a}%
  \BibitemOpen
  \bibfield  {author} {\bibinfo {author} {\bibfnamefont {A.}~\bibnamefont {Zaccone}}, \bibinfo {author} {\bibfnamefont {H.}~\bibnamefont {Wu}}, \ and\ \bibinfo {author} {\bibfnamefont {E.}~\bibnamefont {{Del Gado}}},\ }\bibfield  {title} {\enquote {\bibinfo {title} {{Elasticity of arrested short-ranged attractive colloids: Homogeneous and heterogeneous glasses}},}\ }\href {\doibase 10.1103/PhysRevLett.103.208301} {\bibfield  {journal} {\bibinfo  {journal} {Phys. Rev. Lett.}\ }\textbf {\bibinfo {volume} {103}},\ \bibinfo {pages} {1--5} (\bibinfo {year} {2009})},\ \Eprint {http://arxiv.org/abs/0901.4713} {arXiv:0901.4713} \BibitemShut {NoStop}%
\bibitem [{\citenamefont {Zaccone}\ \emph {et~al.}(2011)\citenamefont {Zaccone}, \citenamefont {Gentili}, \citenamefont {Wu}, \citenamefont {Morbidelli},\ and\ \citenamefont {{Del Gado}}}]{Zaccone2011}%
  \BibitemOpen
  \bibfield  {author} {\bibinfo {author} {\bibfnamefont {A.}~\bibnamefont {Zaccone}}, \bibinfo {author} {\bibfnamefont {D.}~\bibnamefont {Gentili}}, \bibinfo {author} {\bibfnamefont {H.}~\bibnamefont {Wu}}, \bibinfo {author} {\bibfnamefont {M.}~\bibnamefont {Morbidelli}}, \ and\ \bibinfo {author} {\bibfnamefont {E.}~\bibnamefont {{Del Gado}}},\ }\bibfield  {title} {\enquote {\bibinfo {title} {{Shear-driven solidification of dilute colloidal suspensions}},}\ }\href {\doibase 10.1103/PhysRevLett.106.138301} {\bibfield  {journal} {\bibinfo  {journal} {Phys. Rev. Lett.}\ }\textbf {\bibinfo {volume} {106}},\ \bibinfo {pages} {1--6} (\bibinfo {year} {2011})},\ \Eprint {http://arxiv.org/abs/1103.1865} {arXiv:1103.1865} \BibitemShut {NoStop}%
\bibitem [{\citenamefont {Ovarlez}\ \emph {et~al.}(2013)\citenamefont {Ovarlez}, \citenamefont {Tocquer}, \citenamefont {Bertrand},\ and\ \citenamefont {Coussot}}]{Ovarlez2013a}%
  \BibitemOpen
  \bibfield  {author} {\bibinfo {author} {\bibfnamefont {G.}~\bibnamefont {Ovarlez}}, \bibinfo {author} {\bibfnamefont {L.}~\bibnamefont {Tocquer}}, \bibinfo {author} {\bibfnamefont {F.}~\bibnamefont {Bertrand}}, \ and\ \bibinfo {author} {\bibfnamefont {P.}~\bibnamefont {Coussot}},\ }\bibfield  {title} {\enquote {\bibinfo {title} {{Rheopexy and tunable yield stress of carbon black suspensions}},}\ }\href {\doibase 10.1039/c3sm27650c} {\bibfield  {journal} {\bibinfo  {journal} {Soft Matter}\ }\textbf {\bibinfo {volume} {9}},\ \bibinfo {pages} {5540--5549} (\bibinfo {year} {2013})}\BibitemShut {NoStop}%
\bibitem [{\citenamefont {Raney}\ \emph {et~al.}(2018)\citenamefont {Raney}, \citenamefont {Compton}, \citenamefont {Mueller}, \citenamefont {Ober}, \citenamefont {Shea},\ and\ \citenamefont {Lewis}}]{raney2018}%
  \BibitemOpen
  \bibfield  {author} {\bibinfo {author} {\bibfnamefont {J.~R.}\ \bibnamefont {Raney}}, \bibinfo {author} {\bibfnamefont {B.~G.}\ \bibnamefont {Compton}}, \bibinfo {author} {\bibfnamefont {J.}~\bibnamefont {Mueller}}, \bibinfo {author} {\bibfnamefont {T.~J.}\ \bibnamefont {Ober}}, \bibinfo {author} {\bibfnamefont {K.}~\bibnamefont {Shea}}, \ and\ \bibinfo {author} {\bibfnamefont {J.~A.}\ \bibnamefont {Lewis}},\ }\bibfield  {title} {\enquote {\bibinfo {title} {Rotational 3d printing of damage-tolerant composites with programmable mechanics},}\ }\href@noop {} {\bibfield  {journal} {\bibinfo  {journal} {Proceedings of the National Academy of Sciences}\ }\textbf {\bibinfo {volume} {115}},\ \bibinfo {pages} {1198--1203} (\bibinfo {year} {2018})}\BibitemShut {NoStop}%
\bibitem [{\citenamefont {Link}\ \emph {et~al.}(2023)\citenamefont {Link}, \citenamefont {Strybny}, \citenamefont {Divoux}, \citenamefont {Sowoidnich}, \citenamefont {Coenen}, \citenamefont {Gstöhl}, \citenamefont {Schlepütz}, \citenamefont {Zuber}, \citenamefont {Hellmann}, \citenamefont {Rößler}, \citenamefont {Lützenkirchen}, \citenamefont {Heberling}, \citenamefont {Manneville}, \citenamefont {Schäfer}, \citenamefont {Ludwig},\ and\ \citenamefont {Haist}}]{Link2023}%
  \BibitemOpen
  \bibfield  {author} {\bibinfo {author} {\bibfnamefont {J.}~\bibnamefont {Link}}, \bibinfo {author} {\bibfnamefont {B.}~\bibnamefont {Strybny}}, \bibinfo {author} {\bibfnamefont {T.}~\bibnamefont {Divoux}}, \bibinfo {author} {\bibfnamefont {T.}~\bibnamefont {Sowoidnich}}, \bibinfo {author} {\bibfnamefont {M.}~\bibnamefont {Coenen}}, \bibinfo {author} {\bibfnamefont {S.}~\bibnamefont {Gstöhl}}, \bibinfo {author} {\bibfnamefont {C.~M.}\ \bibnamefont {Schlepütz}}, \bibinfo {author} {\bibfnamefont {M.}~\bibnamefont {Zuber}}, \bibinfo {author} {\bibfnamefont {S.}~\bibnamefont {Hellmann}}, \bibinfo {author} {\bibfnamefont {C.}~\bibnamefont {Rößler}}, \bibinfo {author} {\bibfnamefont {J.}~\bibnamefont {Lützenkirchen}}, \bibinfo {author} {\bibfnamefont {F.}~\bibnamefont {Heberling}}, \bibinfo {author} {\bibfnamefont {S.}~\bibnamefont {Manneville}}, \bibinfo {author} {\bibfnamefont {T.}~\bibnamefont {Schäfer}}, \bibinfo {author} {\bibfnamefont {H.-M.}\ \bibnamefont {Ludwig}}, \ and\ \bibinfo {author}
  {\bibfnamefont {M.}~\bibnamefont {Haist}},\ }\bibfield  {title} {\enquote {\bibinfo {title} {Mechanisms of thixotropy in cement suspensions considering influences from shear history and hydration},}\ }\href {\doibase https://doi.org/10.1002/cepa.2810} {\bibfield  {journal} {\bibinfo  {journal} {ce/papers}\ }\textbf {\bibinfo {volume} {6}},\ \bibinfo {pages} {698--704} (\bibinfo {year} {2023})}\BibitemShut {NoStop}%
\bibitem [{\citenamefont {Spicer}, \citenamefont {Hosseini},\ and\ \citenamefont {Babayekhorasani}(2025)}]{Spicer2025}%
  \BibitemOpen
  \bibfield  {author} {\bibinfo {author} {\bibfnamefont {P.~T.}\ \bibnamefont {Spicer}}, \bibinfo {author} {\bibfnamefont {M.}~\bibnamefont {Hosseini}}, \ and\ \bibinfo {author} {\bibfnamefont {F.}~\bibnamefont {Babayekhorasani}},\ }\href {\doibase 10.1016/j.cocis.2025.101916} {\enquote {\bibinfo {title} {Complex fluid product microstructure imaging with light sheet fluorescence microscopy},}\ } (\bibinfo {year} {2025})\BibitemShut {NoStop}%
\bibitem [{\citenamefont {Bonfanti}\ \emph {et~al.}(2020)\citenamefont {Bonfanti}, \citenamefont {Kaplan}, \citenamefont {Charras},\ and\ \citenamefont {Kabla}}]{Bonfanti2020}%
  \BibitemOpen
  \bibfield  {author} {\bibinfo {author} {\bibfnamefont {A.}~\bibnamefont {Bonfanti}}, \bibinfo {author} {\bibfnamefont {J.~L.}\ \bibnamefont {Kaplan}}, \bibinfo {author} {\bibfnamefont {G.}~\bibnamefont {Charras}}, \ and\ \bibinfo {author} {\bibfnamefont {A.}~\bibnamefont {Kabla}},\ }\bibfield  {title} {\enquote {\bibinfo {title} {{Fractional viscoelastic models for power-law materials}},}\ }\href {\doibase 10.1039/d0sm00354a} {\bibfield  {journal} {\bibinfo  {journal} {Soft Matter}\ }\textbf {\bibinfo {volume} {16}},\ \bibinfo {pages} {6002--6020} (\bibinfo {year} {2020})}\BibitemShut {NoStop}%
\bibitem [{\citenamefont {Teixeira}(1988)}]{Teixeira1988}%
  \BibitemOpen
  \bibfield  {author} {\bibinfo {author} {\bibfnamefont {J.}~\bibnamefont {Teixeira}},\ }\bibfield  {title} {\enquote {\bibinfo {title} {{Small-angle scattering by fractal systems}},}\ }\href {\doibase 10.1107/S0021889888000263} {\bibfield  {journal} {\bibinfo  {journal} {J. Appl. Crystallogr.}\ }\textbf {\bibinfo {volume} {21}},\ \bibinfo {pages} {781--785} (\bibinfo {year} {1988})}\BibitemShut {NoStop}%
\bibitem [{\citenamefont {Richards}\ \emph {et~al.}(2017)\citenamefont {Richards}, \citenamefont {Hipp}, \citenamefont {Riley}, \citenamefont {Wagner},\ and\ \citenamefont {Butler}}]{Richards2017}%
  \BibitemOpen
  \bibfield  {author} {\bibinfo {author} {\bibfnamefont {J.~J.}\ \bibnamefont {Richards}}, \bibinfo {author} {\bibfnamefont {J.~B.}\ \bibnamefont {Hipp}}, \bibinfo {author} {\bibfnamefont {J.~K.}\ \bibnamefont {Riley}}, \bibinfo {author} {\bibfnamefont {N.~J.}\ \bibnamefont {Wagner}}, \ and\ \bibinfo {author} {\bibfnamefont {P.~D.}\ \bibnamefont {Butler}},\ }\bibfield  {title} {\enquote {\bibinfo {title} {{Clustering and Percolation in Suspensions of Carbon Black}},}\ }\href {\doibase 10.1021/acs.langmuir.7b02538} {\bibfield  {journal} {\bibinfo  {journal} {Langmuir}\ }\textbf {\bibinfo {volume} {33}},\ \bibinfo {pages} {12260--12266} (\bibinfo {year} {2017})}\BibitemShut {NoStop}%
\end{thebibliography}


\section{Appendix}
\label{sec:app}

\subsection{Rheological measurements}

\begin{figure*}
    \includegraphics[scale=0.5, clip=true, trim=0mm 0mm 0mm 0mm]{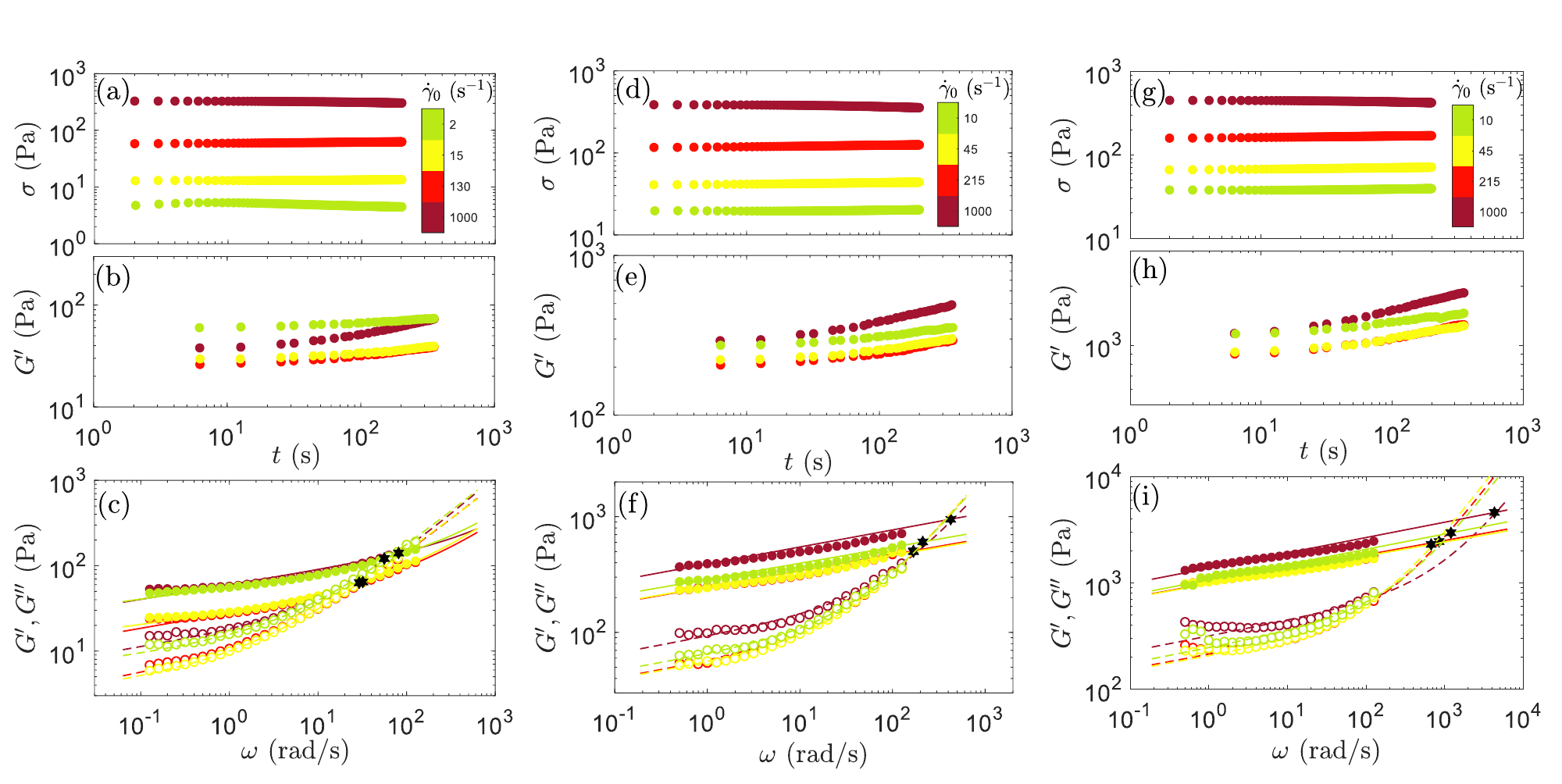}
    \centering
    \caption{Example of results for the pre-shear protocol with a pre-shear time of $200~\mathrm{s}$, corresponding to volume fractions of CB particles of $1.6$ [(a)-(c)], 2.5 [(d)-(f)] and $3.2~\%$ [(g)-(i)]. Each set of panel displays the shear stress vs time during the pre-shear step, the elastic modulus vs time during rest following flow cessation and the viscoelastic spectra of the aged gels. Black markers correspond to the crossover point $(G_c, \omega_c)$, where $G^{\prime}_c(\omega) = G^{\prime\prime}_c(\omega)$.}
    \label{fig:supRheol13}
\end{figure*}

\begin{figure}
    \includegraphics[scale=0.45, clip=true, trim=0mm 0mm 0mm 0mm]{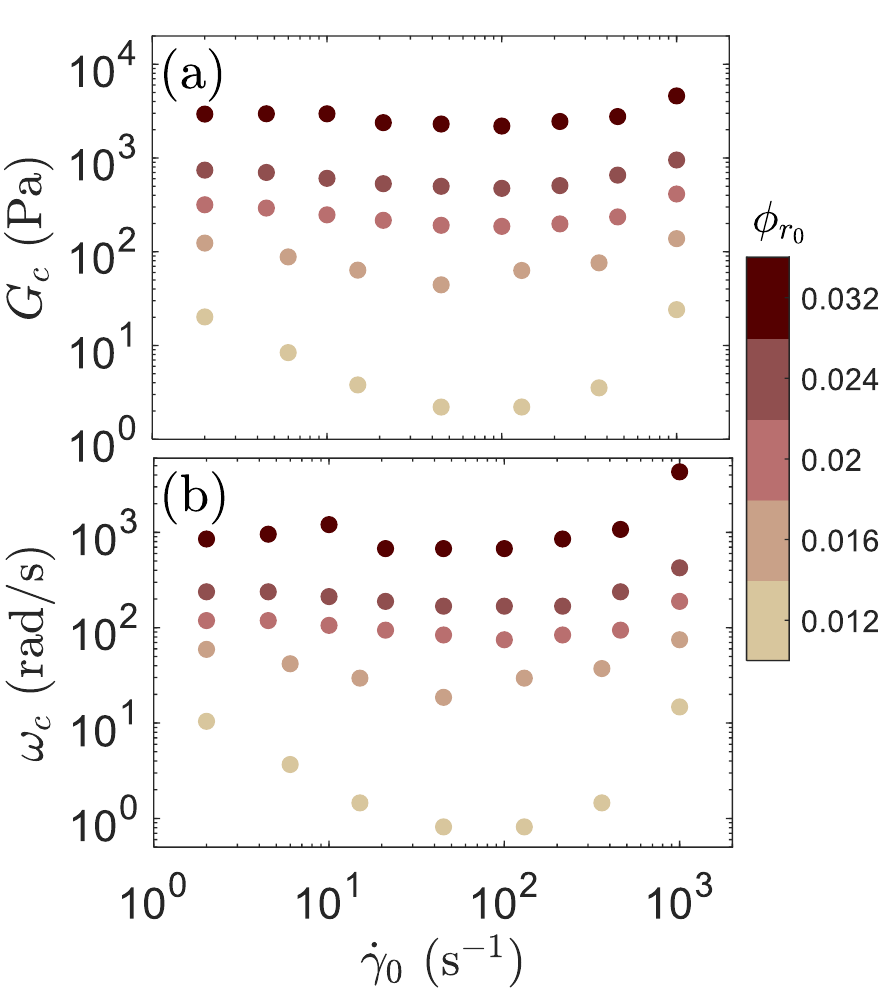}
    \centering
    \caption{Crossover modulus $(G_c)$ and angular frequency $(\omega_c)$ vs pre-shear rate obtained for different volume fractions of CB particles.
    }
    \label{fig:supRheol14}
\end{figure}

\noindent Our main rheological dataset uses a pre-shear time of $200~\mathrm{s}$. For $\dot{\gamma}_0 > 7~\mathrm{s}^{-1}$, this duration is more than sufficient to reach a steady state. Fig.~\ref{fig:supRheol13}(a)–(i) shows the stress evolution during the flow step-down, followed by gel aging and the viscoelastic spectra after flow cessation, for various pre-shear rates and volume fractions. The corresponding crossover modulus ($G_c$) and frequency ($\omega_c$), indicated by black markers in Fig.~\ref{fig:supRheol13}(c), (f), and (i), are reported in Fig.~\ref{fig:supRheol14} as function of the pre-shear rate.

All acquired spectra are well described by the Fractional Kelvin–Voigt model [sketched in Fig.~\ref{fig:rheol}(b)], which enables extraction of the crossover point even when it lies outside the experimental frequency window. The model reads~\cite{Bonfanti2020}:

\begin{equation}
\begin{aligned}
    G^{\prime}(\omega) &= \mathbb{V} \cdot \omega^{\alpha} \cos\left( \frac{\alpha \pi}{2} \right) + \mathbb{G} \cdot \omega^{\beta} \cos\left( \frac{\beta \pi}{2} \right) \\
    G^{\prime\prime}(\omega) &= \mathbb{V} \cdot \omega^{\alpha} \sin\left( \frac{\alpha \pi}{2} \right) + \mathbb{G} \cdot \omega^{\beta} \sin\left( \frac{\beta \pi}{2} \right)
\end{aligned}
\end{equation}

\noindent In this model the crossover coordinates ($\omega_c$, $G_c$) are given by

\begin{equation}
\begin{aligned}
\omega_c
&=
\Biggl(\frac{\mathbb{V}\,\sin\bigl(\tfrac{\alpha\pi}{2}\bigr)}
{\mathbb{G}\,\sin\bigl(\tfrac{\beta\pi}{2}\bigr)}\Biggr)^{\frac{1}{\beta-\alpha}} \\
G_c &=  2\,\mathbb{V}\,\omega_c^{\alpha}\,\sin\bigl(\tfrac{\alpha\pi}{2}\bigr)
= 2\,\mathbb{G}\,\omega_c^{\beta}\,\sin\bigl(\tfrac{\beta\pi}{2}\bigr)
\end{aligned}
\end{equation}

All experimental data could be fitted using $\alpha \approx 0.92$ and $\beta \approx 0.16$.

\begin{figure}[t!]
    \includegraphics[scale=0.5, clip=true, trim=0mm 0mm 0mm 0mm]{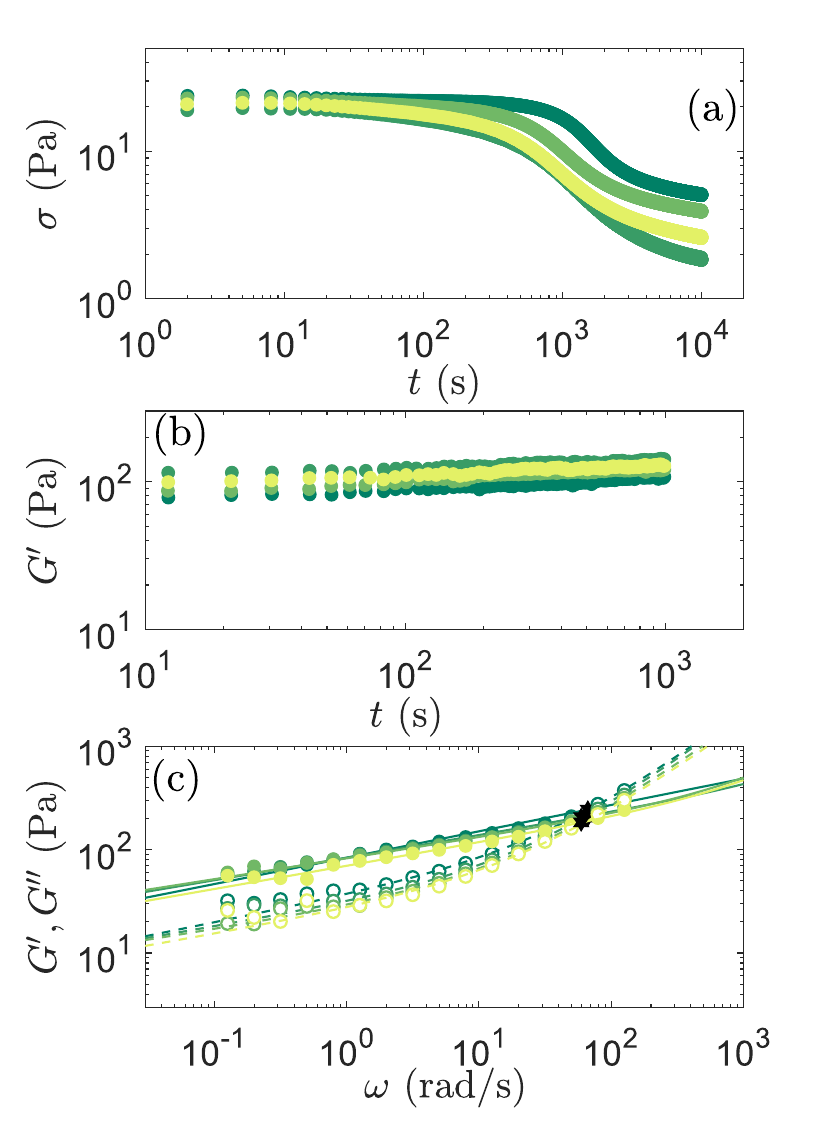}
    \centering
    \caption{Example of results for the pre-shear protocol with a pre-shear time of $10^4~\mathrm{s}$, corresponding to a 3.2~\% volume fraction of CB particles. Color codes for the pre-shear rate. From dark to light green: $\dot{\gamma}_0$ = 3, 2, 1 and $0.5~\mathrm{s}^{-1}$. (a) Shear stress vs time during the pre-shear step. (b) Elastic modulus vs time during rest following flow cessation. (c) Viscoelastic spectra of the aged gels. Black markers correspond to the crossover point $(G_c, \omega_c)$, where $G^{\prime}_c(\omega) = G^{\prime\prime}_c(\omega)$.}
    \label{fig:supRheol6}
\end{figure}

\begin{figure}[t!]
    \includegraphics[scale=0.5, clip=true, trim=0mm 0mm 0mm 0mm]{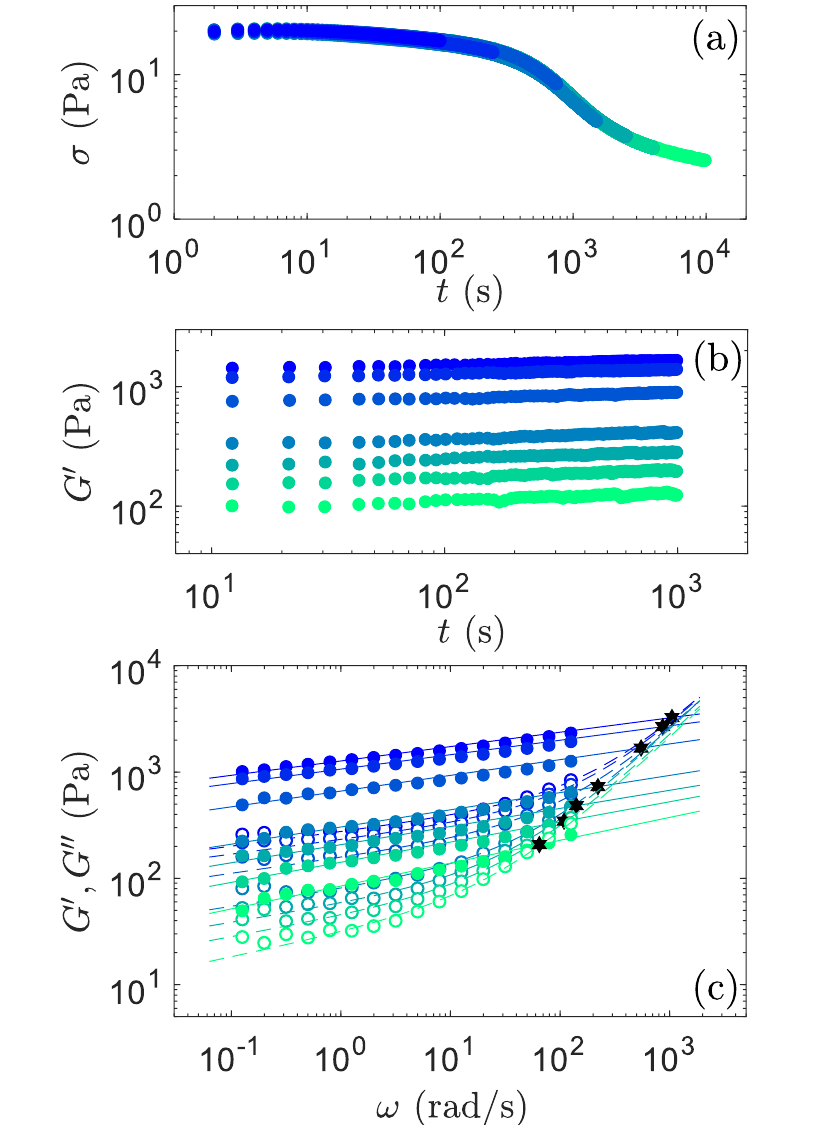}
    \centering
    \caption{Effect of shearing time in the anti-thixotropic regime on the viscoelastic properties of CB gels with a volume fraction of $\phi = 0.032$. (a) Shear stress vs time during the pre-shear step. The color codes for the shearing time. (b) Elastic modulus vs time during rest following flow cessation. (c) Viscoelastic spectra of the aged gels. Black markers correspond to the crossover point $(G_c, \omega_c)$, where $G^{\prime}_c(\omega) = G^{\prime\prime}_c(\omega)$.}
    \label{fig:supRheol5}
\end{figure}

For $\dot{\gamma}_0 \leq 7~\mathrm{s}^{-1}$, the long transient regime—referred to as ``antithixotropy'', requires a significantly longer pre-shear time, on the order of $10^4~\mathrm{s}$. This critical shear rate, which marks the upper limit of the antithixotropic regime in CB dispersions, was thoroughly characterized in our previous work~\cite{Bauland2025}. Fig.~\ref{fig:supRheol6} presents the stress evolution during the step-down protocol at different $\dot{\gamma}_0 < 7~\mathrm{s}^{-1}$, along with the corresponding gel aging and viscoelastic spectra after flow cessation, for a volume fraction of $3.2~\%$. These data correspond to the low elasticity regime (quantified via the crossover modulus $\tilde{G}_c$) resulting from the lowest pre-shear rates shown in Fig.~\ref{fig:rheol}(c) of the main text.
Figure~\ref{fig:supRheol5} shows the time-dependent response of a CB dispersion at a volume fraction of $3.2\%$ sheared at $
\dot{\gamma}_0 = 0.05~\mathrm{s}^{-1}$.
We observe that the stress decreases over time. Moreover, the longer the system is held at $\dot{\gamma}_0 = 0.05~\mathrm{s}^{-1},$ the less elastic the gel formed after flow cessation becomes.

\subsection{USAXS measurements}

\begin{figure}[t!]
    \includegraphics[scale=0.45, clip=true, trim=0mm 0mm 0mm 0mm]{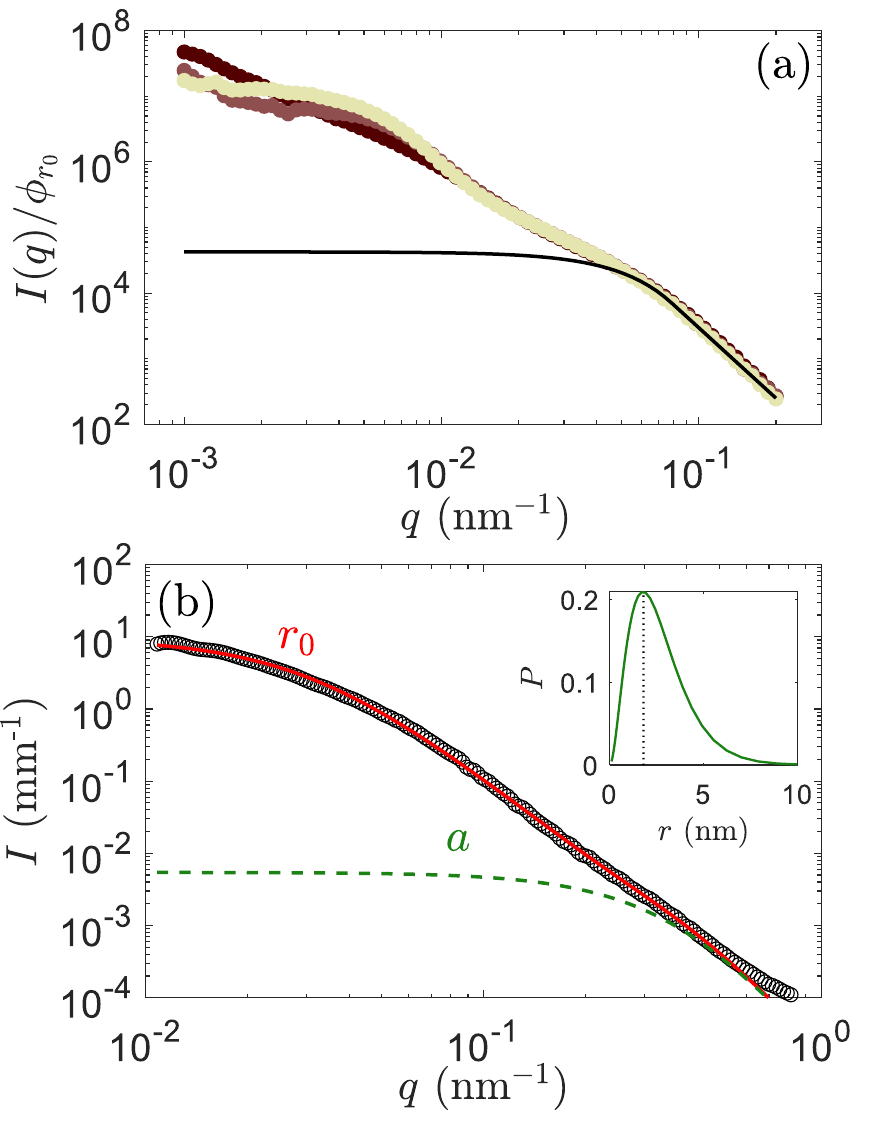}
    \centering
    \caption{(a) Scattering intensity $I(q)$ vs $q$ of CB dispersions under shear ($\dot{\gamma}=10^3~\rm s^{-1}$) rescaled by the volume fraction of particles $\phi_{r_0} =0.6$, $1.2$ and $3.2~\%$. Black line represents the fit of the primary particles contribution by a Guinier-Porod model. (b) Form factor of the CB particles measured by SAXS for $\phi_{r_0}$ = 10$^{-4}$. CB primary particles are composed of nodules of radius $a$ that are fused to form primary aggregates of radius $r_0$. Inset: log-normal distribution of $a$. }
    \label{fig:supSAXS2}
\end{figure}

The structure of CB dispersions, as probed by USAXS, is modeled using a hierarchical framework described in~\cite{Hipp2021,Bouthier2022b}, which accounts for the organization of the primary CB particles ($r_0$) into small fractal clusters ($\xi_1$), which further aggregate into a larger-scale fractal network ($\xi_2$). Each level is described by a mass fractal structure factor $S_i$~\cite{Teixeira1988}. As previously reported, the primary particles themselves are composed of small ``nodules'' ($a$) fused together to form the primary particles of radius $r_0$~\cite{Richards2017,Hipp2021}.\jb{Note that previous studies have referred to the nodules as the primary particles of CB dispersions~\cite{Richards2017}. Here, we adopt a simplified definition by designating the primary particles as the smallest indivisible structural unit, namely $r_0$.} The form factor of the CB particle is thus described as the product of the form factor of a sphere $P(q)$ and a mass fractal structure factor $S_0(q)$. The full model reads:

\begin{equation}
    I\left(q\right)= \phi_a V_a(\Delta\rho)^2 
    \underbrace{P(q) 
    S_0(q)}_{\text{$r_0$}} \underbrace{S_1(q)}_{\text{$\xi_1$}} \underbrace{S_2(q)}_{\text{$\xi_2$}}
\end{equation}

with
\begin{equation}
\begin{array}{l}
\left\{
\begin{array}{ll}
\text{ } & a < r_0 < \xi_1 < \xi_2 \\
P(q) & = \bigg[\frac{3[\sin(qa) - qa\cos(qa)]}{(qa)^3} \bigg]^2 \\
S_i(q) & = 1 + \frac{d_{f_i} \Gamma(d_{f_i} - 1)}{[1 + 1/(q\xi_i)^2]^{(d_{f_i} - 1)/2}} \cdot \frac{\sin[(d_{f_i} - 1)\tan^{-1}(q\xi_i)]}{(q R_{i})^{d_{f_i}}}
\end{array}
\right.
\end{array}.
\label{eq:SAXS}
\end{equation}

\noindent where $\phi_a$ and $V_a$ denote the volume fraction and unit volume of the nodules of size $a$, respectively, and $\Delta \rho$ is the scattering length density difference between mineral oil and CB particles. \jb{In Eq.~\ref{eq:SAXS}, $\xi_i$ denotes $a$, $r_0$, and $\xi_1$ for $S_0(q)$, $S_1(q)$, and $S_2(q)$, respectively. $R_i$ represents a cutoff length, smaller than the corresponding cluster size $\xi_i$ but larger than the maximum size captured in $S_{i-1}(q)$. We treated $R_i$ as a fitting parameter and found that it typically lies between $2\xi_{i-1}$ and $4\xi_{i-1}$.}

The radius of gyration $R_{g_i}$ associated with the characteristic scattering length $\xi_i$ is calculated following~\cite{Hipp2021}:
\begin{equation}
    R_{g_i}^2 = \frac{d_{f_i}(d_{f_i} + 1)\xi_i^2}{2}
\end{equation}

\begin{figure}[t!]
    \includegraphics[scale=0.45, clip=true, trim=0mm 0mm 0mm 0mm]{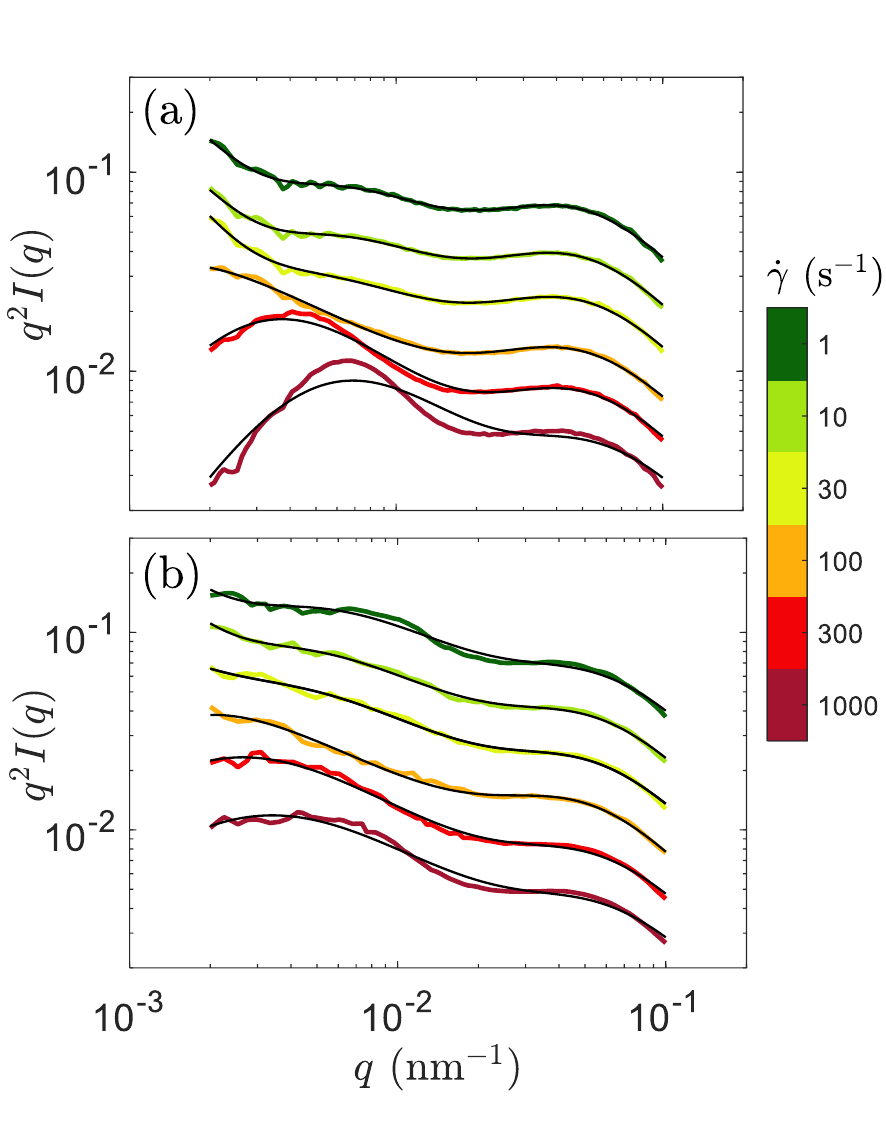}
    \centering
    \caption{Scattering intensity $q^2I(q)$ vs $q$ of the $1.6~\rm \%$ CB dispersion during (a) flow and (b) after flow cessation fitted with the hierarchical fractal model.}
    \label{fig:supSAXS1}
\end{figure}

We now describe the determination of fixed parameters in the model. From a dilute CB dispersion, we measured the form factor of CB particles [Fig.~\ref{fig:supSAXS1}(b)] and found $a = 1.8~\rm nm$, $d_{f_0} = 2.85$, and $r_0 = 76~\rm nm$~\cite{Bauland2024}.

The volume fraction of nodules, $\phi_a$, is calculated from the mass fraction of CB particles using: 
\[
\phi_a = \frac{c_w}{c_w + \frac{d_{\rm cb}}{d_{\rm oil}}(1 - c_w)},
\]
where $d_{\rm cb} = 2.26$ and $d_{\rm oil} = 0.871$ denote the densities of CB and oil, respectively. Given the high fractal dimension of primary particles ($d_{f_0} \approx 3$), indicating that the CB particles behave essentially as non‑fractal spheres with rough surfaces, we approximate $\phi_a \approx \phi_{r_0}$. This assumption is supported by the successful rescaling of scattering intensity using $\phi_{r_0}$ across various volume fractions [Fig.~\ref{fig:supSAXS1}(a)].

The scattering length density difference, $\Delta \rho$, is estimated by analyzing $I(q)$ at high $q$ for different CB volume fractions. Considering only the scattering from primary particles, one has $\lim_{q \to \infty} I(q)/\phi_{r_0} = V_{r_0}(\Delta \rho)^2$ [Eq.~\ref{eq:SAXS}]. Using the intensity plateau and particle size obtained from a Guinier–Porod fit [Fig.~\ref{fig:supSAXS1}(a)], we determine $R_g = 46~\rm nm$, in good agreement with $r_0$ from the form factor. This yields $\Delta \rho = 4.8 \times 10^8~\rm mm^{-2}$, close to the previously reported value of $4.4 \times 10^8~\rm mm^{-2}$ for CB in hydrogenated propylene carbonate~\cite{Richards2017}.

Fig.~\ref{fig:supSAXS2}(a)–(b) shows fits of the scattering curves using the hierarchical model. The parameters associated with the primary particle level are fixed across all fits ($r_0 = 76~\rm nm$, $d_{f_0} = 2.9$, $a = 1~\rm nm$), as well as $\phi_a = 0.016$ and $\Delta \rho = 4.8 \times 10^8~\rm mm^{-2}$.

The fitting of higher structural levels involves three or six free parameters, depending on whether a third level is included. The cutoff length of the fractal regime, $R_i$, is taken as approximately 2–3 times the size of the preceding unit $\xi_{i-1}$, consistent with previous reports~\cite{Hipp2021}.

In the hierarchical model of CB gels, particle mass conservation reads:

\begin{equation}
\begin{array}{l}
\left\{
\begin{array}{ll}
\rho = \frac{\phi_{r_0}}{V_{r_0}} \\
\rho = \frac{\left(\frac{\xi_2}{\xi_1}\right)^{d_{f_2}} \left(\frac{\xi_1}{r_0}\right)^{d_{f_1}}}{\xi_2^3}
\end{array}
\right.
\end{array}
\label{eq:mass}
\end{equation}
\begin{figure}[t!]
    \includegraphics[scale=0.5, clip=true, trim=0mm 0mm 0mm 0mm]{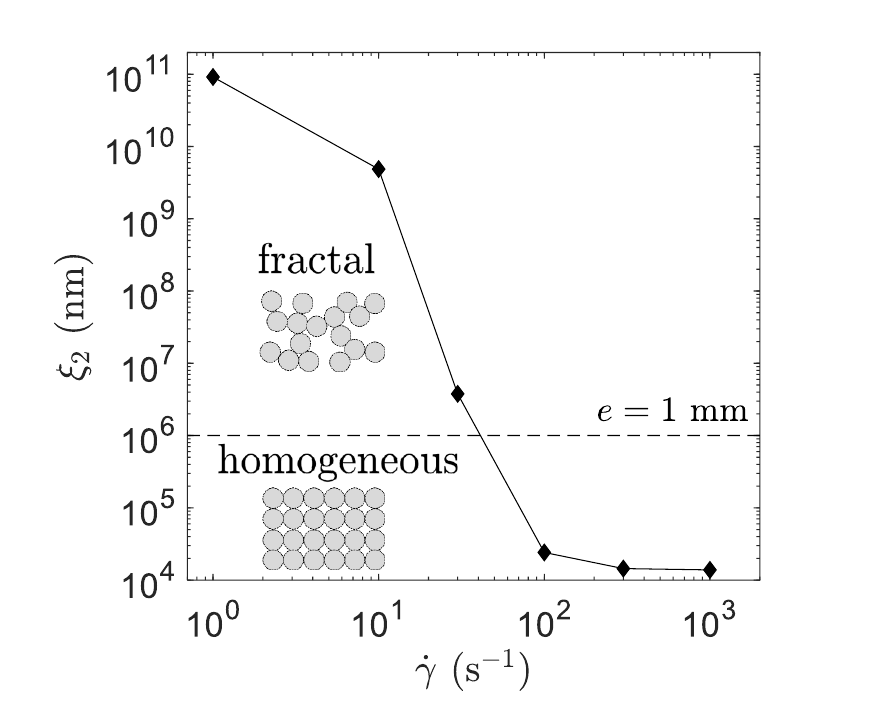}
    \centering
    \caption{Mesh size $\xi_2$ of the large-scale fractal network calculated from mass conservation using the scattering data of the $1.6~\%$ CB gels. The gap size of the rheometer $e = 1~\rm mm$ is indicated by a black dotted line for comparison.}
    \label{fig:supSAXS3}
\end{figure}
\noindent where $\rho$ is the number density of primary particles. As an example, for the $1.6~\%$ dispersion, using $r_0 = 76~\rm nm$, we calculate $\rho = 2.7 \times 10^{10}~\rm mm^{-3}$. For gel states, $\xi_2$ is computed from Eq.~\ref{eq:mass} and plotted as a function of pre-shear rate in Fig.~\ref{fig:supSAXS3}. For $\dot{\gamma}_0 < 100~\rm s^{-1}$, the derived $\xi_2$ exceeds the geometry gap, which is unphysical. This observation suggests that the gel cannot be described as randomly packed aggregates, but rather forms a heterogeneous network at larger scales~\cite{Bouthier2022b}.

\subsection{Estimation of the CB interaction parameters}

\begin{figure}[t!]
    \includegraphics[scale=0.35, clip=true, trim=0mm 0mm 0mm 0mm]{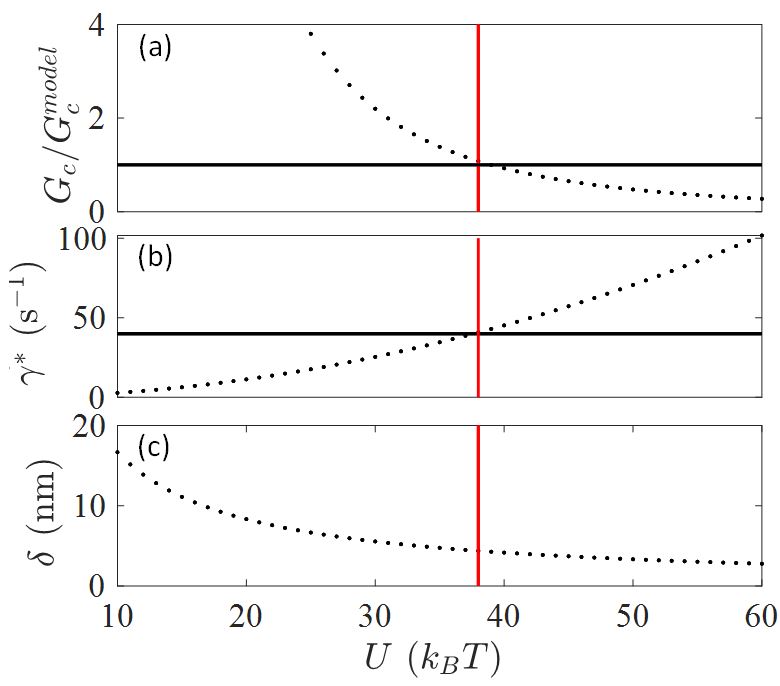}
    \centering
    \caption{Optimization of the CB interaction parameters with respect to our experimental data using $U$ as an input. 
(a) The dotted line represents $G_c/G_c^{\mathrm{model}}$, where $G_c^{\mathrm{model}}$ is calculated from Eq.~\ref{eq:G}. 
The horizontal black line indicates perfect agreement between the experimental data and the model, i.e., $G_c/G_c^{\mathrm{model}} = 1$. 
(b) The dotted line represents the critical shear rate calculated for $Mn = 1$, given by $\dot{\gamma} = (U/\delta)/(6\pi\eta R_{\mathrm{g1}}^2)$. 
The horizontal black line corresponds to the experimental critical shear rate, $\dot{\gamma}^* = 40~\mathrm{s}^{-1}$. 
(c) Dot line represents $\delta = A a / (12U)$ as a function of $U$. 
The red vertical line marks the optimal potential value, $U = 38~k_B T$, corresponding to $\delta = 4.4~\mathrm{nm}$, based on the experimental constraints from the elastic moduli and the critical shear rate.
}
    \label{fig:supU}
\end{figure}

\tg{To connect the van der Waals interaction to the square-well model used in our colloidal gel analysis, we use the relation $A = \frac{12 U \delta}{a}$
which provides a convenient mapping between a generic short-range attraction and an equivalent van der Waals interaction between colloidal particles. Here, $A$ is the Hamaker constant, $a$ the particle radius, and $U$ and $\delta$ denote the effective depth and range of the square-well potential, respectively. 
}

\tg{This relation is derived from the Derjaguin approximation applied to the London dispersion interaction. 
For two spheres of radius $a$ separated by a distance $h \ll a$, the non-retarded van der Waals interaction is 
$U_{\rm vdW}(h) \simeq - \frac{A a}{12 h}$.
To map the real van der Waals interaction onto an effective square--well potential, we require that the integrated cohesive energy over the attractive range be the same in both descriptions: $\int_0^{\delta} \frac{A a}{12 h}  dh \approx U$.
However, $\int_0^{\delta} \frac{1}{h} \, dh = \ln \left( \frac{\delta} h_{\min} \right),$ which diverges as $h_{\min} \to 0$. A standard approximation in colloid physics (used, for example, in Baxter's sticky--sphere theory and in mappings of short--range potentials) is to replace the logarithmic divergence by an effective range $\delta$, such that $\frac{A a}{12} \frac{1}{\delta} \sim U$. Solving for $A$ gives $A = \frac{12 U \delta}{a}.$}

\tg{Previous studies of carbon-black particles dispersed in mineral oil report $U \approx 30~k_B T$~\cite{Varga2019, Trappe2007}. Dagastine~\cite{dagastine2002} calculated the Hamaker constant for carbon black in tetradecane, obtaining $A = 1.15\times 10^{-19}~\mathrm{J}$; correcting for the refractive index of our mineral oil ($n=1.47$) yields $A = 1.09\times 10^{-19}~\mathrm{J}$. Using this corrected value of $A$ with $U = 30~k_B T$ gives $\delta = \frac{A a}{12 U} = 5.6~\mathrm{nm}$.}

\tg{Rather than using $U \approx 30~k_B T$~\cite{Varga2019, Trappe2007}, which was obtained for a different type of carbon dispersed in a different mineral oil, we optimized $U$ and, consequently, $\delta$ based on our experimental data, namely the critical shear rate $\dot{\gamma}^* = 40~\mathrm{s}^{-1}$ and $G_c$. 
The best fits are obtained for $U = 38~k_B T$ and $\delta = 4.4~\mathrm{nm}$ as shown in~Fig.\ref{fig:supU}. These optimized values are used throughout the main body of the article.
We note that the optimized $U$ is 25\% higher than the estimation reported in~\cite{Varga2019, Trappe2007}, which is reasonable given the differences in CB particle type and solvent.
 }

\end{document}